\documentstyle[12pt,epsf]{article}
\def\largelinestretch{\renewcommand{\baselinestretch}{1.3}}
\largelinestretch\small\normalsize

\title{
\vspace*{-10mm}
\hfill
\parbox{4cm}{\large JINR E2-99-236}\\
\vspace*{10mm}
\bf Phenomenological analysis of $\varepsilon^{'}/\varepsilon$
    within an effective chiral lagrangian approach\\
    at $O(p^6)$}
 \author{
A.A.Bel'kov${}^1$\thanks{E-mail: {\tt belkov@cv.jinr.dubna.su}}~,
G.Bohm${}^2$\thanks{E-mail: {\tt bohm@ifh.de}}~,
A.V.Lanyov${}^1$\thanks{E-mail: {\tt lanyov@mail.desy.de}}~,
A.A.Moshkin${}^1$\thanks{E-mail: {\tt andrem@cv.jinr.dubna.su}}
\\[1ex]
\small
${}^1$
        Particle Physics Laboratory, Joint Institute for Nuclear
        Research,
\hfill\\[-0.2ex]
\small
        141980 Dubna, Moscow region, Russia
\hfill\\[0.2ex]
\small
${}^2$
DESY -- Zeuthen, Platanenallee 6, D-15735 Zeuthen, Germany
\hfill\\
}
\date{}
\begin{document}

%\largelinestretch\normalsize
\thispagestyle{empty}
\begin{titlepage}
\thispagestyle{empty}
\maketitle
\begin{abstract}
%\normalsize
    We have combined a new systematic calculation of mesonic matrix
elements at $O(p^6)$ from an effective chiral lagrangian approach with
Wilson coefficients taken from \cite{buras3}, derived in the framework
of perturbative QCD, and restricted partly by experimental data.
    We derive complete expressions for $K\to 2\pi$ amplitudes and compare 
the results for $\varepsilon^{'}/\varepsilon$ with experiment.
\end{abstract}

\end{titlepage}

%============================================================================
\section{Introduction}
%============================================================================

    The starting point for most calculations of nonleptonic
kaon decays is an effective weak lagrangian of the form
\cite{vzsh,gilman-wise}
\begin{equation}
{\cal L}^{q}_{w}\big(|\Delta S|=1\big) =
\sqrt2\,G_F\,V_{ud}V^{*}_{us}\sum_{i} C_i \, {\cal O}_i
\label{weak-lagr}
\end{equation}
which can be derived with the help of the Wilson operator product
expansion from elementary quark processes, with additional gluon
exchanges.
   In the framework of perturbative QCD the coefficients $C_i$ are
to be understood as scale and renormalization scheme dependent
functions.
   There exist extensive next-to-leading order (NLO) calculations
\cite{buras1,ciuchini1} in the context of kaon decays, among others.
   These calculations are based on the possibility of factorization of
short- and long-distance contributions into Wilson coefficient
functions $C_i$ and mesonic matrix elements of four-quark operators
${\cal O}_i$, respectively.
   The latter, however, can presently be obtained only by using
nonperturbative, i.e. model-dependent, methods, because not only perturbative
QCD breaks down at scales $\mu \le 1 \mbox{GeV}$, but also the QCD degrees of 
freedom (quarks and gluons) have to be replaced by the mesonic ones.
   Thus, a fully satisfactory solution would include the theoretical 
understanding of confinement. 
   The only consistent approach to this problem may be found in lattice 
calculations. 
   A discussion of the present status has been given in \cite{lattice} and 
will not be repeated here.

   Usually, the results of calculations are displayed with the help of
$B$-factors in the form
\begin{equation}
T_{K\to 2\pi} = \sqrt2\,G_F\,V_{ud}V^{*}_{us} \sum_{i}
                 \Big[ C_i(\mu)B_i(\mu) \Big]
                 <\pi\pi |{\cal O}_i|K>_{vac.sat.}\,,
\label{B_definition}
\end{equation}
where the mesonic matrix elements of four-quark operators are approximated 
by their vacuum saturation values, which are real and $\mu$-independent.
   In principle, factors $B_i(\mu)$ should be estimated by some 
higher-order calculations in the long-distance regime, for instance, in
$1/N_c$-expansion \cite{buras2} in the form $1+O(1/N_c)$, or from the
lattice approach.
   The preliminary stage of these calculations is best characterized
by the long standing difficulties to explain quantitatively the well-known 
$\Delta I = 1/2$ rule.
   Of course, the lack of such calculations for long-distance effects
severely restricts the predictive power of (\ref{weak-lagr}), leaving
only the possibility for some semi-phenomenological treatment
\cite{buras1,buras3,buras4}, with correspondingly large theoretical
uncertainties.

   The main aim of the present paper is a further semiphenomenological
treatment of the long-distance (non-perturbative) aspects of the
above lagrangian, especially in view of the actuality of the task to
analyze the implications of the measured parameter of direct CP violation, 
$\varepsilon^{'}/\varepsilon$. 
   The features distinguishing our approach from others 
\cite{buras1,ciuchini1,ciuchini2,bertolini1,bertolini2,paschos1} are mainly 
the following:\\
-- At first, a chiral lagrangian up to $O(p^8)$ is used for deriving mesonic 
currents and densities, from which the matrix elements up to $O(p^6)$
are constructed.\\
-- At second, according to Weinberg's power counting scheme \cite{weinberg},
the calculation includes tree level, one- and two-loop diagrams, whereby the 
renormalization of the perturbative (loop) expansion for the matrix elements 
in question makes use of superpropagator regularization (to be discussed 
below), being connected to the intrinsic scale $4\pi F_0$ of the chiral 
lagrangian, and showing good stability (decreasing higher order 
contributions).\\
   While the consistency of this approach could be checked 
phenomenologically in the strong interaction (hadronic) sector, in the 
application to the weak $\Delta S = 1$ interactions there remains the above 
problem of matching the scale- and scheme dependence of the $C_i$, thereby
bridging the gap between short and long distance treatments. 

   In the present paper, we perform the calculation of matrix elements
successively for increasing orders in $p^2$, displaying also the intermediate
results, in order to analyze the trend of the successive expansion
terms. 
   An obstacle to this procedure is the proliferation of structure
constants in higher order chiral lagrangians, which have to be fixed
by experiment. 
   This has been accomplished up to now only to $O(p^4)$ 
\cite{gasser1,gasser2,dafne}.
   As a way out, we invoke another effective model -- the
Nambu-Jona-Lasinio (NJL) model \cite{njl} -- whose modifications have been 
used by several groups \cite{ebert,bijnens1,p6-our} to ``derive''
the chiral lagrangian by bosonization of the fermionic degrees of
freedom, suitably adapting the free parameters to reproduce those
of the chiral lagrangian. 
   In this framework, the structure constants of higher order lagrangians 
can be calculated, and they are well comparable in the $O(p^4)$ case with the 
empirical ones. 
   Therefore it seems justified, to estimate effects of orders beyond 
$O(p^4)$ by taking NJL-derived structure constants.

   If we compare the predicted amplitudes (\ref{B_definition}) with 
experiment, it turns out that even after replacing 
$<\pi\pi |{\cal O}_i|K>_{vac.sat.}$ by higher order matrix
elements there are still some correction factors needed, which
we call $\widetilde{B}_i$.
   We restrict ourselves in this paper to the display of their ranges and 
correlations, especially taking into account the large value $\varepsilon^{'}$
obtained in the NA31 experiment \cite{NA31-CP} and confirmed recently by the 
KTeV \cite{KTeV-CP} and NA48 \cite{NA48-CP} collaborations.
   As our approach to the renormalization of chiral perturbation theory
involves no arbitrary cut-off or scale other than $F_0$, the bare $\pi /K$
decay constant, there is also no other possibility to match the scale
dependence of the Wilson coefficients, except that due to the
renormalization of $F_0$ (and the other bare parameters of the
effective lagrangian). 
   This is at least partly included in our approach, as we redefine the bare
coupling $F_0$ for each order to reach agreement with $\pi /K \to \mu \nu$ 
decay. 
   As mentioned above, this procedure is stable and consistent, i.e. does not
lead to large higher order corrections or large renormalization effects.

   In section 2 we repeat all relevant definitions taken from our earlier 
work. 
   Section 3 discusses the higher order structure constants used in the 
calculation of $K\to \pi\pi$ amplitudes, the latter being sketched in 
section 4. 
   The last two sections give our results and conclusions.   

%============================================================================
\section{Lagrangians and currents}
%============================================================================

   In the present paper we use the operators ${\cal O}_i$
in the representation given in \cite{vzsh,bijnens-wise}:
\begin{eqnarray*}
{\cal O}_1 &=&   \bar{u}_L \gamma_\mu u_L \; \bar{d}_L \gamma^\mu s_L
       -   \bar{d}_L \gamma_\mu u_L \; \bar{u}_L \gamma^\mu s_L\,,
\nonumber\\
{\cal O}_2 &=&   \bar{u}_L \gamma_\mu u_L \; \bar{d}_L \gamma^\mu s_L
       +   \bar{d}_L \gamma_\mu u_L \; \bar{u}_L \gamma^\mu s_L
       + 2 \bar{d}_L \gamma_\mu d_L \; \bar{d}_L \gamma^\mu s_L
       + 2 \bar{s}_L \gamma_\mu s_L \; \bar{d}_L \gamma^\mu s_L\,,
\nonumber\\
{\cal O}_3 &=&   \bar{u}_L \gamma_\mu u_L \; \bar{d}_L \gamma^\mu s_L
       +   \bar{d}_L \gamma_\mu u_L \; \bar{u}_L \gamma^\mu s_L
       + 2 \bar{d}_L \gamma_\mu d_L \; \bar{d}_L \gamma^\mu s_L
       - 3 \bar{s}_L \gamma_\mu s_L \; \bar{d}_L \gamma^\mu s_L\,,
\nonumber\\
{\cal O}_4 &=&   \bar{u}_L \gamma_\mu u_L \; \bar{d}_L \gamma^\mu s_L
       +   \bar{d}_L \gamma_\mu u_L \; \bar{u}_L \gamma^\mu s_L
       -   \bar{d}_L \gamma_\mu d_L \; \bar{d}_L \gamma^\mu s_L\,,
\nonumber\\
{\cal O}_5 &=& \bar{d}_L \gamma_\mu \lambda^a_c s_L
  \left(\sum_{q=u,d,s}\bar{q}_R\,\gamma^\mu\,\lambda^a_c\,q_R\right)\,,
\quad
{\cal O}_6 = \bar{d}_L \gamma_\mu s_L
  \left( \sum_{q=u,d,s} \bar{q}_R \, \gamma^\mu \, q_R \right)\,,
\nonumber\\
{\cal O}_7 &=& 6\bar{d}_L \gamma_\mu s_L
  \left( \sum_{q=u,d,s} \bar{q}_R \, \gamma^\mu \, Q \, q_R \right)\,,
\quad
{\cal O}_8 = 6\bar{d}_L \gamma_\mu \lambda^a_c s_L
  \left(\sum_{q=u,d,s}\bar{q}_R\,\gamma^\mu\,\lambda^a_c\,Q\,q_R\right)\,,
\end{eqnarray*}
where $q_{L,R} = \frac12 \, (1\mp\gamma_5) q$; $\lambda^a_c$ are
the generators of the $SU(N_c)$ color group; $Q$ is the matrix of
electric quark charges.
   The operators ${\cal O}_{5,6}$ containing right-handed
currents are generated by gluonic penguin diagrams and the analogous
operators ${\cal O}_{7,8}$ arise from electromagnetic
penguin diagrams.
   The operators ${\cal O}_{1,2,3,5,6}$ and ${\cal O}_4$ describe the
transitions with $\Delta I = 1/2$ and $\Delta I = 3/2$, respectively,
while the operators ${\cal O}_{7,8}$ contribute to the transition with both
$\Delta I = 1/2$ and $\Delta I = 3/2$.

   The Wilson coefficients $C_i$ of the effective weak lagrangian
(\ref{weak-lagr}) with four-quark operators ${\cal O}_i$
are connected with the Wilson coefficients $c_i$ corresponding to the
basis of four-quark operators $Q_{i}$ given in Refs.
\cite{buras3,buras1}, by the following linear relations:
\begin{eqnarray}
&&C_1 = c_1-c_2+c_3-c_4+c_9-c_{10}\,,\quad
  C_2 = \frac{1}{5}\big(c_1+c_2-c_9-c_{10}\big)+c_3+c_4\,,
\nonumber \\&&
  C_3 = \frac{1}{5} C_4
      = \frac{1}{5}\bigg(
        \frac{2}{3}\big(c_1+c_2\big)+c_9+c_{10}\bigg)\,,
\nonumber \\&&
  C_5 = c_6\,,\quad C_6 = 2\bigg(c_5+\frac{1}{3}c_6\bigg)\,,\quad
  C_7 = \frac{1}{2}\bigg(c_7+2c_8\bigg)\,,\quad C_8 = \frac{1}{4}c_8\,.
\label{ci-param}
\end{eqnarray}

   The bosonized version of the effective Lagrangian (\ref{weak-lagr}) can be
expressed in the form \cite{bos-weak}:
 %%%%%%%%%%
 \def\jim{(J^1_{L\mu}\! - i J^2_{L\mu})}
 \def\rim{(J^1_R\! - i J^2_R)}
 \def\jiii{(J^3_{L\mu}\!+\! \frac1{\sqrt3} J^8_{L\mu})}
 \def\riii{(J^3_R\! -\!\frac1{\sqrt3} J^8_R\! - \sqrt\frac23 \, J^0_R)}
 \def\jivp{(J^4_{L\mu}\! + i J^5_{L\mu})}
 \def\livp{(J^4_L\! + i J^5_L)}
 \def\jvip{(J^6_{L\mu}\! + i J^7_{L\mu})}
 \def\lvip{(J^6_L\! + i J^7_L)}
 \def\rvip{(J^6_R\! + i J^7_R)}
 %%%%%%%%%%
\begin{eqnarray}
 {\cal L}^{mes}_{w} &=& \widetilde{G}_F\Bigg\{
  (-\xi_1\! + \xi_2\! + \xi_3) \bigg[ \jim \jivp - \jiii \jvip \bigg]
\nonumber\\
&&+ (\xi_1 + 5\,\xi_2) \sqrt\frac23 \, J^0_{L\mu} \jvip
+ {10 \over \sqrt3} \, \xi_3 \, J^8_{L\mu} \jvip
\nonumber\\
&&+ \xi_4 \bigg[ \jim \jivp + 2 \, J^3_{L\mu} \jvip \bigg]
\nonumber\\
&&- 4 \, \xi_5 \bigg[ \rim \livp - \riii \lvip
\nonumber\\
&&\qquad - \sqrt\frac23 \, \rvip (\sqrt2 J^8_L - J^0_L)
           \bigg]
\nonumber\\
&&+ \xi_6 \, \sqrt\frac32 \, \jivp J^0_{R\mu}
+ 6 \, \xi_7 \, \jvip (J^3_{R\mu} + \frac1{\sqrt3} \, J^8_{R\mu})
\nonumber\\
&&- 16 \, \xi_8 \, \bigg[ \rim \livp + \frac12 \, \riii \lvip
\nonumber\\
&&\qquad + \frac1{\sqrt6} \, \rvip (\sqrt2 \, J^8_L - J^0_L)
            \bigg]
 \Bigg\} + \mbox{h.c.}
\label{weak-mes}
\end{eqnarray}
   Here $\widetilde{G}_F = \sqrt2\,G_F\,V_{ud}V^{*}_{us}$, 
$J^a_{L/R \, \mu}$ and $J^a_{L/R}$ are bosonized $(V\mp A)$ and $(S\mp P)$ 
meson currents and densities, corresponding to the quark currents
$\bar q \gamma_\mu \frac14 (1\mp \gamma^5) \lambda^a q$ and densities
$\bar q \frac14 (1\mp \gamma^5) \lambda^a q$, respectively ($\lambda^a$ are
the generators of the $U(3)_F$ flavor group);
\begin{eqnarray}
&&\xi_1 = C_1 \left( 1 - {1 \over N_c} \right)\,, \qquad
\xi_{2,3,4} = C_{2,3,4} \left( 1 + {1 \over N_c} \right)\,,
\nonumber\\
&&\xi_{5,8} = C_{5,8}\left( 1 - {1 \over N_c^2} \right) + {1 \over 2 N_c} C_{6,7}\,, \qquad
\xi_{6,7} = C_{6,7}\,,
\label{xi-param}
\end{eqnarray}
where the color factors ${1/N_c}$ originate from Fierz-transformations of
four-quark operators ${\cal O}_i$ (see more technical details in 
\cite{bos-weak}).

   Only the even-intrinsic-parity sector of the chiral strong lagrangian is 
required to describe nonleptonic kaon decays up to and including $O(p^6)$.
   The meson currents/densities $J^a_{L/R\mu}$ and $J^a_{L/R}$ are obtained 
from the quark determinant by variation over additional external sources
associated with corresponding quark currents and densities
\cite{bos-weak}.
   From the momentum expansion of the quark determinant to $O(p^{2n})$
one can derive the strong lagrangian for mesons ${\cal L}_{eff}$ of
the same order and the corresponding currents and densities $J^a_{L/R\mu}$ 
and $J^a_{L/R}$ to the order $O(p^{2n-1})$ and $O(p^{2n-2})$, respectively.
   For example, from the terms of quark determinant of $O(p^2)$ one obtains
the following:
\begin{eqnarray}
&&{\cal L}^{(p^2)}_{eff}=-\frac{F^2_0}{4} \,\mbox{tr}\,
                      \big( L^2_{\mu}\big)
                     +\frac{F_0^2}{4} \,\mbox{tr}\,
                      \big( \chi U^\dagger + U\chi^\dagger \big)\,,
\nonumber \\&&
   J^{(p^1)a}_{L\mu} = \frac{iF^2_0}{4}\,\mbox{tr}\,(\lambda^a L_\mu)\,,
\quad
   J^{(p^0)a}_{L} =
   \frac{F^2_0}{4}\overline{m}R\,\,\mbox{tr}\,(\lambda^a U)\,,
\label{l2}
\end{eqnarray}
where $U = \exp \left(\frac{i\sqrt{2}}{F_0} \varphi \right)$,
with $\varphi$ being the pseudoscalar meson matrix,
and $L_{\mu}=D_{\mu}U\,U^\dagger$,
$D_{\mu}U = \partial_{\mu}U + (A^L_{\mu}U - U A^R_{\mu})$ and
$A^{R/L}_\mu = V_\mu \pm A_\mu$ are right/left-handed combinations of
vector and axial-vector fields.
   Furthermore, $F_0 \approx 90$ MeV is the bare coupling constant of pion
decay, $\chi =\mbox{diag}\big(\chi^2_u,\chi^2_d,\chi^2_s\big)
= -2m_0\!\!<\!\!\overline{q}q\!\!>\!\!F_0^{-2}$ is the meson mass
matrix, $\chi^2_u = 0.0114\,\mbox{GeV}^2$, $\chi^2_d = 0.025\,\mbox{GeV}^2$, 
$\chi^2_s = 0.47\,\mbox{GeV}^2$, $m_0$ is the current quark mass matrix,
$<\!\!\overline{q}q\!\!> = (- 220 \mbox{MeV})^3$ is the quark condensate, 
$\overline{m} \approx 265$ MeV is an average constituent quark mass, and
$R = <\!\!\overline{q}q\!\!>\!\!/(\overline{m}F_0^2) = -4.96$.

   At $O(p^4)$ one gets
\begin{eqnarray}
{\cal L}^{(p^4)}_{eff}&\Rightarrow&
    \bigg(L_1-\frac{1}{2}L_2 \bigg)\,\big( \,\mbox{tr}\, L^2_{\mu}\big)^2
  + L_2 \,\mbox{tr}\, \bigg(\frac{1}{2}[L_{\mu},L_{\nu}]^2
  + 3(L^2_{\mu})^2\bigg)
  + L_3 \,\mbox{tr}\, \big[ (L^2_{\mu})^2 \big]
\nonumber \\&&
  - L_4 \,\mbox{tr}\, \big( L^2_{\mu}\big)\,
        \,\mbox{tr}\, \big( \chi U^\dagger + U \chi^\dagger \big)
  - L_5 \,\mbox{tr}\,\Big( L^2_{\mu}
          \big( \chi U^\dagger + U \chi^\dagger \big)\Big)
\nonumber \\&&
  + L_8 \,\mbox{tr}\, \Big( (\chi ^\dagger U)^2 + (\chi U^\dagger)^2 \Big)
  + H_2 \,\mbox{tr}\, \chi \chi ^\dagger \,\,,
\nonumber \\
   J^{(p^3)a}_{L\mu} &\Rightarrow& i\,\mbox{tr}\,\bigg\{\lambda^a\bigg[
    L_4 L_\mu\,\mbox{tr}(\chi U^\dagger +U\chi^\dagger )
  + \frac{1}{2}L_5 \{ L_\mu ,(\chi U^\dagger +U\chi^\dagger ) \}\bigg]
    \bigg\}\,,
\nonumber \\
   J^{(p^2)a}_{L} &\Rightarrow& -\overline{m}R\,\mbox{tr}\,
                                 \Big\{\lambda^a \Big[
    L_4 U\,\mbox{tr}(L^2_\mu)
   +L_5 (L^2_\mu U) -2L_8 U\chi^\dagger U -H_2\chi \Big]\Big\}\,,
\label{l4}
\end{eqnarray}
where $L_i$ and $H_2$ are structure constants introduced by Gasser and
Leutwyler \cite{gasser1}.

   For the sake of brevity, here and in the following expressions for the 
lagrangian at $O(p^6)$ we restrict ourselves to the terms which are 
necessary to calculate the decay $K\to 2\pi$.
   At $O(p^6)$ one needs the following terms:
\footnote{The rather lengthy full expression for the bosonized effective
          lagrangian at $O(p^6)$ was presented in Refs.\cite{p6-our}.}:
\begin{eqnarray}
{\cal L}_{eff}^{(p^6)} &\Rightarrow& \mbox{tr} \bigg\{
          Q_{12} \Big(\,\chi R^\mu U^\dagger \big(
            D_\mu D_\nu U + D_\nu D_\mu U \big) U^\dagger L^\nu
\nonumber \\ && \quad \quad
           +\chi^\dagger L^\mu U \big(
            \overline{D}_\mu \overline{D}_\nu U^\dagger
           +\overline{D}_\nu \overline{D}_\mu U^\dagger \big) UR^\nu
          \Big)
\nonumber \\ &&
         +Q_{13}\Big[
            \chi (  \overline{D}_\mu\overline{D}_\nu U^\dagger
                  L^\mu L^\nu
              + R^\nu R^\mu U
                \overline{D}_\mu \overline{D}_\nu U^\dagger )
\nonumber \\ && \quad \quad
           +\chi^\dagger (D_\mu D_\nu U R^\mu R^\nu
                         +L^\nu L^\mu D_\mu D_\nu U) \Big]
\nonumber \\ &&
         +Q_{14}\Big[{\chi}\Big(U^{\dagger}D_{\mu}D_{\nu}U\overline{D}^{\mu}
          \overline{D}^{\nu}U^\dagger+\overline{D}_{\mu}\overline{D}_{\nu}
          U^{\dagger}D^{\mu}D^{\nu}UU^{\dagger}\Big)
\nonumber \\ && \quad \quad
                 +{\chi}^{\dagger}\Big(U\overline{D}_{\mu}
          \overline{D}_{\nu}U^{\dagger}D^{\mu}D^{\nu}U
          +D_{\mu}D_{\nu}U\overline{D}^{\mu}
          \overline{D}^{\nu}U^{\dagger}U\Big)\Big]
\nonumber \\ &&
         +Q_{15} \chi^\dagger L_{\mu} \chi R^{\mu}
         +Q_{16}\Big( \chi^\dagger \chi R_{\mu} R^{\mu}
              +\chi \chi^\dagger L_{\mu} L^{\mu} \Big)
\nonumber \\ &&
         +Q_{17}\Big( U\chi^\dagger U \chi^\dagger L_{\mu}
              L^{\mu}+U^\dagger \chi  U^\dagger \chi R_{\mu} R^{\mu} \Big)
         +Q_{18}\Big[(\chi U^\dagger L_{\mu})^2
         + (\chi^\dagger U R_{\mu} )^2 \Big]
\nonumber \\ &&
         +Q_{19}\Big[(\chi  U^\dagger)^3
             + (\chi^\dagger U)^3 \Big]
         +Q_{20}\Big( U^\dagger \chi  \chi^\dagger \chi
                  +U \chi^\dagger \chi  \chi^\dagger \Big)
                         \bigg\}\,,
\label{l6}
\end{eqnarray}
where $Q_i$ are structure constants introduced in \cite{p6-our},
whereas $R_\mu =U^\dagger D_\mu U$.
   The corresponding terms of $(V\mp A)$ and $(S\mp P)$ bosonized
meson currents are given by
\begin{eqnarray*}
J^{(p^5)a}_{L \mu} &\Rightarrow&i\frac{1}{4} \mbox{tr} \bigg\{ \lambda^a \bigg[
 -2 Q_{14} \Big[
                (U\chi^{\dagger}+\chi U^{\dagger})D_{\mu}D_{\nu}U\,
                 U^{\dagger}L^{\nu}
               +D_{\mu}D_{\nu}U(U^{\dagger}\chi U^{\dagger}+\chi ^{\dagger})
                                   L^{\nu}
\nonumber \\&&~~~~~~~~~
               -U\overline{D}^{\nu}\Big(
                       (U^{\dagger}\chi +\chi ^{\dagger}U)\overline{D}_{\nu}
                                 \overline{D}_{\mu}U^{\dagger}
                        +\overline{D}_{\nu}\overline{D}_{\mu}U^{\dagger}
                          (U\chi ^{\dagger}+\chi U^{\dagger})
                                    \Big)
\nonumber \\&&~~~~~~~~~
               +L^{\nu}U\Big(
                       (U^{\dagger}\chi +\chi ^{\dagger}U)\overline{D}_{\mu}
                                 \overline{D}_{\nu}U^{\dagger}
                        +\overline{D}_{\mu}\overline{D}_{\nu}U^{\dagger}
                           (U\chi ^{\dagger}+\chi U^{\dagger})
                        \Big)
\nonumber \\&&~~~~~~~~~
               +D^{\nu}\Big((U\chi ^{\dagger}+\chi U^{\dagger})D_{\nu}D_{\mu}U
                        +D_{\nu}D_{\mu}U(U^{\dagger}\chi +\chi ^{\dagger}U)
                       \Big)U^{\dagger}
           \Big]
\nonumber \\&&
 +2 Q_{15} \big(U{\chi}^{\dagger}L_{\mu}{\chi}U^{\dagger}
                 +{\chi}U^{\dagger}L_{\mu}U{\chi}^{\dagger}
            \big)
 +2 Q_{16} \big(\big\{U{\chi}^{\dagger}{\chi}U^{\dagger},
                               L_{\mu}\big\}
                 +\big\{{\chi}{\chi}^{\dagger},L_{\mu}\big\}
           \big)
\nonumber \\&&
 +2 Q_{17} \big(\big\{\big(U{\chi}^{\dagger}\big)^2,L_{\mu}\big\}
                +\big\{\big({\chi}U^{\dagger}\big)^2,L_{\mu}\big\}
           \big)
\nonumber \\&&
 -4 Q_{18} \big( U{\chi}^{\dagger}L_{\mu}U{\chi}^{\dagger}
                +{\chi}U^{\dagger}L_{\mu}{\chi}U^{\dagger}
           \big)
           \bigg]\bigg\}\,,
\label{curv5}
\end{eqnarray*}
and
\begin{eqnarray*}
J^{(p^4)a}_L &\Rightarrow& \overline{m}R\mbox{tr}\Big\{ \lambda^a \Big[
  Q_{12}\, L^{\mu} U\{\overline{D}_{\mu},\overline{D}_{\nu}\}U^\dagger
           \, UR^{\nu}
\nonumber \\&&
 +Q_{13} \big( L^{\nu}L^{\mu}D_{\mu}D_{\nu}U
              +D_{\mu}D_{\nu}U\cdot R^{\mu}R^{\nu}\big)
\nonumber \\&&
 +Q_{14} \big( U\overline{D}^{\nu}\overline{D}^{\mu}U^{\dagger}
                             D_{\nu}D_{\mu}U
                            +D_{\nu}D_{\mu}U
                \overline{D}^{\nu}\overline{D}^{\mu}U^{\dagger}\cdot U\big)
\nonumber \\&&
 +Q_{15}\, L^{\mu} \chi R_{\mu}
 +Q_{16}\,\big( \chi R_{\mu}^2 +L_{\mu}^2 \chi \big)
 +Q_{17}  \big( U \chi^{\dagger} UR_{\mu}^2
          +L_{\mu}^2 U \chi^{\dagger} U \big)
\nonumber \\&&
 +2Q_{18}\, L^{\mu}U \chi^{\dagger} L_{\mu}U
 + Q_{19} \big(U\chi ^{\dagger}\big)^2U
 +Q_{20} (\chi U^{\dagger}\chi +\chi \chi ^{\dagger}U +U\chi ^{\dagger}\chi )
           \Big] \Big\}\,.
\label{curs4}
\end{eqnarray*}

   We do not show explicitly the terms of the effective action at $O(p^8)$
generating the scalar current $J^{(p^6)}_L$ which is necessary for the
full calculation of the tree-level matrix elements at $O(p^6)$ for the
penguin operators, since the corresponding contributions turn out to
be negligibly small.

%============================================================================
\section{Structure constants}
%============================================================================

    For numerical estimates of $K\to2\pi$ amplitudes and 
$\varepsilon^{'}/\varepsilon$ we will need the values of the structure
constants $L_i$ and $Q_i$ which were introduced in the effective chiral 
lagrangians at $O(p^4)$ and $O(p^6)$, respectively. 
    The current experimental status of the effective chiral lagrangian
at $O(p^4)$ has been discussed within ChPT in some detail in \cite{dafne}.
    For the $O(p^4)$ lagrangian (\ref{l4}) all structure constants $L_i$ 
are at present determined phenomenologically as measurable values $L_i^r$ 
depending on the renormalization scale $\tilde{\mu}$.
    The best values of the parameters $L_i$ quoted at a $\rho$-meson mass 
scale and the sources of the experimental information used are listed in 
table 1. 
    The scale dependence of the measurable coefficients $L_i^r$ is determined
by relation
\begin{equation}
  L^r_i(\mu_2) = L^r_i(\mu_1) + \frac{\Gamma_i}{(4\pi)^2} 
                                \mbox{ln} \frac{\mu_1}{\mu_2}\,,
\label{scale}
\end{equation}
where the coefficients $\Gamma_i$ are also given in table 1.

   In the context of the scale dependence of the structure coefficients
$L_i$, we have to note that in our approach the UV divergences resulting from 
meson loops at $O(p^4)$ and $O(p^6)$ were separated by using the 
superpropagator (SP) regularization method \cite{sp-volkov} which 
particularly well suits the treatment of loops in nonlinear chiral 
theories.
   The result is related to the dimensional regularization technique though 
some the difference lies in the scale parameter $\tilde{\mu}$ which is no 
longer arbitrary but fixed by the inherent scale of the chiral theory
$\tilde{\mu}=4\pi F_0\approx 1$ GeV, and the UV divergences have to be
replaced by a finite term using the following substitution:
$$
\big({\cal C}-1/\varepsilon\big) \to C_{SP} = -1+4{\cal C} +\beta \pi\,,
$$
where ${\cal C} = 0.577$ is Euler's constant,
$\varepsilon = (4-D)/2$, and $\beta$ is an arbitrary constant
introduced by the Sommerfeld-Watson integral representation of the
superpropagator based on unitarity.

   The phenomenological analysis of the so-called Skyrme and non-Skyrme 
structures in the effective chiral lagrangian at $O(p^4)$ was earlier
carried out in \cite{belkov-old} by using the direct SP-calculations of meson 
loops for $\pi\pi$-scattering amplitudes.
   After reformulating this analysis in terms of the structure coefficients 
$L_i$, the values
\begin{equation}
L_1 = ( 0.6 \pm 0.2)\cdot 10^{-3}\,,\quad
L_2 = ( 1.6 \pm 0.3)\cdot 10^{-3}\,,\quad
L_3 = (-3.5 \pm 0.6)\cdot 10^{-3}
\label{pipi-SP}
\end{equation}
were obtained from the experimental data on $\pi\pi$-scattering lengths.
   In the same way, taking into account the tadpole loops, the splitting of 
the decay constants $F_\pi$ and $F_K$ was used at $O(p^4)$ to fix 
$C_{SP}\approx 3.0$ and $L_5 = (1.6 \pm 0.3)\cdot 10^{-3}$.
   The latter value as well as the  values (\ref{pipi-SP}) are in a good
agreement with the corresponding ones given in table 1.
   This fact indicates that the choice $\tilde{\mu}= m_{\rho}$ for the 
renormalization scale of ChPT proves to be consistent with the internal scale
of SP-regularization. 
   Therefore, we use the values of $L_i$ given in table 1 for further
phenomenological analysis.

    The structure constants $Q_i$ of the $O(p^6)$ lagrangian (\ref{l6}) are
still not defined from experiment.
    Therefore we need some theoretical model to estimate their values.
    Both the structure constants $L_i$ and $Q_i$ can be obtained from the 
modulus of the logarithm of the quark determinant of the NJL-type model 
\cite{njl} which explicitly contains, apart from the pseudoscalar Goldstone 
bosons, also scalar, vector and axial-vector resonances as dynamic degrees of 
freedom.
    However, in order to avoid double counting in calculating pseudoscalar 
meson amplitudes when taking into account resonance degrees of freedom, one 
has to integrate out (reduce) these resonances in the generating functional
of the bosonization approach.
    As a consequence of this procedure, the structure coefficients of
pseudoscalar low-energy interactions will be quite strongly modified.
In this way one effectively takes into account resonance-exchange
contributions \cite{bijnens1,reduction,p6-our}.

    Without reduction of resonance degrees of freedom, the structure
constants $L_i = N_c/(16 \pi^2)\cdot l_i$, and
$Q_i = N_c/(32 \pi^2 \overline{m}^2)\cdot q_i$ are fixed from the bosonization
of an NJL-type model as
\begin{eqnarray*}
&& l_1 = \frac{1}{2}l_2 = \frac{1}{24}\,,\quad
   l_3=-\frac{1}{6}\,,\quad
   l_4= 0 \,,\quad
   l_5= xy-x\,,
\nonumber \\
&& l_8= \frac{1}{2}xy-x^2y-\frac{1}{24}\,,
\label{lhcoeff}
\end{eqnarray*}
and
\begin{eqnarray*}
&&
q_{12}= \frac{1}{60}\,,\quad
q_{13}= -\frac{1}{3}\bigg(\frac{1}{20}-x+c\bigg)\,,\quad
q_{14}= \frac{x}{6}\,,
\nonumber \\&&
q_{15}= \frac{2}{3}x\big(1-x\big)-
        \bigg(\frac{1}{3}-2x\bigg)c\,,\quad
q_{16}= -\frac{1}{120}+\frac{4}{3}x^2+\frac{x}{6}\big(1-4x\big)
        -2\bigg( x-\frac{1}{6} \bigg)c\,,
\nonumber \\&&
q_{17}=  \frac{1}{120}+\frac{x}{6}\big(1-4x\big)
        -\bigg(x+\frac{1}{6}\bigg)c\,,\quad
q_{18}= \frac{4}{3}x^2+\bigg(\frac{1}{6}-x\bigg)c\,,
\nonumber \\ &&
q_{19}=-\frac{1}{240}-x^2+\frac{2}{3}x^3
       +x\big(1+2xy\big)c\,,
\nonumber \\ &&
q_{20}= \frac{1}{240}+x^2+2\big(1-2y\big)x^3-x\big(1+2xy\big)c\,,
\label{qcoeff}
\end{eqnarray*}
where $x = -\overline{m} F_0^2/(2\!\!<\!\!\overline{q} q \! \! >) = 0.1$,
$y = 4\pi^2F_0^2/(N_c\overline{m}^2)=1.5$ and $c=1-1/(6y)$.

    After reduction of the resonances, the structure coefficients
get the form
\begin{eqnarray*}
&&\!\!\!\!\!\!\!\!\!\!
l^{red}_1 =\frac{1}{2}l^{red}_2
          = \frac{1}{12}\bigg[ Z^8_A
          +2(Z^4_A-1)\bigg(\frac{1}{4}\tilde{y}(Z^4_A-1)-Z^4_A\bigg)\bigg]\,,
\nonumber \\ &&\!\!\!\!\!\!\!\!\!\!
l^{red}_3 = -\frac{1}{6}\bigg[ Z^8_A
          +3(Z^4_A-1)\bigg(\frac{1}{4}\tilde{y}(Z^4_A-1)-Z^4_A\bigg)\bigg]\,,
\nonumber \\ &&\!\!\!\!\!\!\!\!\!\!
l^{red}_4= 0 \,,\quad
l^{red}_5 = (\tilde{y}-1) \frac14 Z^6_A \,,\quad
l^{red}_8 = \frac{\tilde{y}}{16} Z^4_A \,,\quad
h^{red}_2 = \tilde{y} Z^2_A \bigg(\frac{Z^2_A}{8}-x\bigg)\,.
\label{lhred}
\end{eqnarray*}
and
\begin{eqnarray*}
&& \!\!\!\!\!\!\!\!\!\!
q^{red}_{12} = q^{red}_{13}= 0 \,,\quad
q^{red}_{14} = \frac{1}{24}Z^6_A\,,
\nonumber \\ &&\!\!\!\!\!\!\!\!\!\!
q^{red}_{16} = q^{red}_{17} =
-\frac{Z^6_A}{64}\bigg\{\tilde{y}\!
       -Z^2_A\bigg[4\!-6\Big(1+4(1\!-Z^2_A)\Big)(1\!-\tilde{y})
       +4\Big(1+16(1\!-Z^2_A)\Big)
         \frac{1\!-\tilde{y}}{\tilde{y}}\bigg]\bigg\}\,,
\nonumber \\ &&\!\!\!\!\!\!\!\!\!\!
q^{red}_{15} = -2q^{red}_{18} =
\frac{1}{48}Z^6_A\bigg[3\tilde{y}
           -2Z^2_A\bigg(5-12(1-Z^2_A)
           \frac{(1-\tilde{y})^2}{\tilde{y}}\bigg)\bigg]\,,
\nonumber \\ &&\!\!\!\!\!\!\!\!\!\!
q^{red}_{19} = \frac{1}{3}q^{red}_{18} =
-\frac{1}{192}Z^6_A(3\tilde{y}-2)\,,
\label{qred}
\end{eqnarray*}
where $\tilde{y} = 4\pi^2F_0^2/(Z^2_A N_c \overline{m}^2)=2.4$, and
$Z^2_A=0.62$ is the $\pi -A_1$ mixing factor.

    In table 1 we also present the predictions of the NJL model for the 
structure coefficients $L_i$ which after reduction of meson resonances turn
out to be in a good agreement with phenomenology.
    This fact indicates that the NJL-model is a reasonable low-energy
approximation for the effective four-quark interaction, generating a 
realistic effective meson lagrangian. 
    Therefore we also use it to fix the values of the structure constants
$Q_i$ for numerical estimates of the contributions of the $O(p^6)$
lagrangian (\ref{l6}).

%============================================================================
\section{Amplitudes of $K\to 2\pi$ decays}
%============================================================================

   Using isospin relations, the $K \to 2\pi$ decay amplitudes can be
parameterized  as
\begin{eqnarray*}
T_{K^+\to\pi^+\pi^0} &=& {\sqrt3\over 2}\,A_2\,,
\nonumber \\
T_{K^0_S\to\pi^+\pi^-} &=& \sqrt{2\over 3}\,A_0+{1\over\sqrt3}\,A_2\,,
\qquad
T_{K^0_S\to\pi^0\pi^0} = \sqrt{2\over 3}\,A_0-{2\over\sqrt3}\,A_2\,.
\end{eqnarray*}
   The isotopic amplitudes $A_{2,0}$ determine the $K \to 2\pi$
transitions into states with isospin $I=2,0$, respectively:
$$
A_2 = a_2 \, e^{i\delta_2}\,, \qquad A_0 = a_0 \, e^{i\delta_0}\,,
$$
where $\delta_{2,0}$ are the phases of $\pi\pi$-scattering.
   It is well known that direct $CP$ violation results in an additional
(small) relative phase between $a_2$ and $a_0$.
   Let us next introduce the contributions of the four-quark
operators ${\cal O}_i$ to the isotopic amplitudes ${\cal A}_I^{(i)}$
by the relations
\begin{eqnarray}
A_I&=&{\cal F}_I {\cal A}_I\,,\quad
{\cal A}_I = -i\,\sum_{i=1}^8\xi_i {\cal A}_I^{(i)}\,,
\label{defa}
\end{eqnarray}
where
${\cal F}_2 =\sqrt2{\cal F}_0
            = {\sqrt3\over2}\widetilde{G}_F F_0(m_K^2-m_\pi^2)$.

   At $O(p^2)$, corresponding to the soft-pion limit, for the nonzero 
tree-level amplitudes ${\cal A}_I^{(i)}$ we obtain the following 
expressions:
\begin{eqnarray}
&&\!\!\!\!\!\!\!\!\!\!
{\cal A}_0^{(1)}=-{\cal A}_0^{(2,3)}=-{\cal A}_0^{(4)}=-1\,,\quad
{\cal A}_0^{(7)} = -{\cal A}_2^{(7)} = 2\,,\quad
{\cal A}_0^{(5)} = -32\bigg(\frac{R\overline{m}}{F_0}\bigg)^2 L_5 \,,
\nonumber \\&&\!\!\!\!\!\!\!\!\!\!
{\cal A}_0^{(8)} = \frac{16(R\overline{m})^2}{m^2_K-m^2_\pi}
\bigg\{1-\frac{2}{F^2_0}\bigg[ 6L_4(\chi^2_s+\chi^2_d+\chi^2_u)
\nonumber \\&&\quad\quad\quad\quad\quad\quad\quad\quad\quad\quad
        + (L_5-4L_8)(\chi^2_s+3\chi^2_d+2\chi^2_u)
        + 2L_5 m_\pi^2 \bigg]\bigg\}\,,
\nonumber \\&&\!\!\!\!\!\!\!\!\!\!
{\cal A}_2^{(8)} = \frac{8(R\overline{m})^2}{m^2_K-m^2_\pi}
\bigg\{1-\frac{2}{F^2_0}\bigg[ 6 L_4(\chi^2_s+\chi^2_d+\chi^2_u)
\nonumber \\&&\quad\quad\quad\quad\quad\quad\quad\quad\quad\quad
        + (L_5-4L_8)(\chi^2_s+3\chi^2_d+2\chi^2_u)
        + 2L_5 m_K^2 \bigg]\bigg\}\,.
\label{A02-p2}
\end{eqnarray}
   The $L_8$ and $H_2$ contributions in the penguin operators
${\cal O}_{5,8}$ also have a tadpole contribution from
$K \to (\mbox{vacuum})$, included through strong rescattering,
$K\to \pi\pi K$ with $K \to (\mbox{vacuum})$.
   At $O(p^2)$, in case of the penguin operator ${\cal O}_5$, the $L_8$
and $H_2$ contributions to the direct matrix element from $K\to 2\pi$
vertices, are fully canceled by the tadpole diagrams
\footnote{We thank W.A.~Bardeen and A.J.~Buras for drawing our attention
          to this point.}.
   This is due to the possibility of absorbing the tadpole contribution
into a redefinition of the $K\to 2\pi$ vertex if all particles are on
mass shell.
   Moreover, such a cancelation is expected at all orders of $K\to 2\pi$ 
amplitudes including loop diagrams due to general counter term arguments 
given in \cite{gasser2}.
   According to these arguments the structure constant $H_2$ is not directly
measurable and does not occur in the amplitudes of physical processes.

   Some interesting observations on the difference of the momentum behavior
of penguin and non-penguin operators can be drawn from power-counting
arguments.
   According to Eq.\ (\ref{l2}) the leading contributions to the vector 
currents and scalar densities are of $O(p^1)$ and $O(p^0)$, respectively.
   Since in our approach the non-penguin operators are constructed out of
the products of $(V-A)$-currents $J^a_{L\mu}$, while the penguin operators
are products of $(S-P)$-densities $J^a_L$, the lowest-order contributions
of non-penguin and penguin operators are of $O(p^2)$ and $O(p^0)$,
respectively.
   However, due to the well-known cancelation of the contribution of the
gluonic penguin operator ${\cal O}_5$ at lowest order \cite{chivukula}, the 
leading gluonic penguin as well as non-penguin contributions start from 
$O(p^2)$
\footnote{There is no cancellation of the contribution of the
          electromagnetic penguin operator ${\cal O}_8$ at the lowest
          order and the first terms in the expressions (\ref{A02-p2})
          for ${\cal A}_{0,2}^{(8)}$ correspond to the contributions
          at $O(p^0)$.}.
   Consequently, in order to derive the $(V-A)$-currents which contribute 
to the non-penguin transition operators at leading order, it is sufficient 
to use the terms of the quark determinant to $O(p^2)$ only.
   At the same time the terms of the quark determinant to $O(p^4)$ have to 
be kept for calculating the penguin contribution at $O(p^2)$, since it
arises from the combination of $(S-P)$-densities from Eqs.\ (\ref{l2}) and
(\ref{l4}), which are of $O(p^0)$ and $O(p^2)$, respectively.
   In this subtle way a difference in momentum behavior is revealed
between matrix elements for these two types of weak transition operators;
it manifests itself more drastically in higher-order lagrangians and
currents.
   This fact makes penguins especially sensitive to higher order effects.

   Our calculations involve Born and one- and two-loop meson diagrams and 
take into account isotopic symmetry breaking ($\pi^0 - \eta -\eta^{'}$ 
mixing).
   The use of a specialized analytical computation package based on REDUCE
\cite{REDUCE} to calculate amplitudes and loop integration makes
it possible to evaluate a large number of loop diagrams arising for different 
charge channels. 
   The main problem in the calculation of amplitudes at $O(p^6)$ is the 
evaluation of two-loop diagrams.
   A part of them is shown schematically in figure 1 (we do not show rather
trivial diagrams with tadpole loops).
   The diagrams of figure 1a were calculated analytically, because the 
integration in every loop can be performed independently when using the 
superpropagator regularization.
   The two-loop diagrams of figure 1b,c,d cannot be calculated analytically,
but they can be estimated numerically through a dispersion-relation
approach in the same way as it was already done in \cite{bellucci} for the 
so-called ``box'' and ``acnode'' diagrams.
   Such numerical estimates have shown that the contributions of diagrams of 
1b,c,d do not exceed 2\% and can be neglected.

   Table 2 presents the modification of the amplitudes ${\cal A}_I^{(i)}$ 
when including successively the higher order corrections at $O(p^4)$ and 
$O(p^6)$.
   In our numerical estimates the Born contribution at $O(p^4)$ and the 
one-loop contribution at $O(p^6)$ were calculated for central values of 
the phenomenological parameters $L_i$ from table 1.
   The Born contribution at $O(p^6)$ has been estimated for values of
structure constants $Q_i$ fixed from the bosonization of the NJL-model with
reduction of meson resonances. 
   Table 2 shows that the Born contribution at $O(p^6)$ is very small as 
compared to loop contributions and does not play an essential role
in our further analysis of decay amplitudes and $\varepsilon^{'}/\varepsilon$.

   A strong indication that the development to higher orders is physically 
sensitive is given by the behaviour of phases: the strong interaction phases 
$\delta_{2,0}$ arise first at $O(p^4)$, but for the quantitative description 
of the phases it is necessary to go beyond $O(p^4)$.
   At $O(p^4)$, for the $\pi\pi$-scattering phase shifts and their
difference $\Delta=\delta_0-\delta_2$, we have obtained the values
of $\delta_0 \approx 22^\circ$, $\delta_2 \approx -13^\circ$,
$\Delta \approx 35^\circ$ which are in agreement with \cite{kambor}.
   At $O(p^6)$, however, we  obtained $\delta_0 \approx 35^\circ$,
$\delta_2 \approx -9^\circ$, $\Delta \approx 44^\circ$, in a better
agreement with the experimental value $\Delta^{exp} = (48 \pm 4)^\circ$
\cite{belkov-kostyukhin}.

%============================================================================
\section{Phenomenological results}
%============================================================================

   In our approach the parameters $\xi_i$ in Eq.~(\ref{defa}) are
treated as phenomenological ($\mu$-independent) parameters to be fixed
from the experimental data.
   They can be related to the $\mu$-dependent QCD predicted $\xi_i(\mu)$ 
by using some $\mu$-dependent $\widetilde{B}_i$-factor defined as
$$
   \xi_i^{ph}=\xi_i(\mu)\widetilde{B}_i(\mu).
$$
   The factors $\widetilde{B}_i(\mu)$ can be related to the factors 
$B_i(\mu)$ defined in (\ref{B_definition}) by obvious relations.
   Table 3 shows the QCD predictions for the coefficients
$\xi_i(\mu) = \xi_i^{(z)}(\mu)+\tau \xi_i^{(y)}(\mu)$ which correspond
to the Wilson coefficients
$$
  c_i(\mu) = z_i(\mu)+\tau y_i(\mu),\quad
  \tau = -\frac{V_{td}V_{ts}^{*}}{V_{ud}V_{us}^{*}}\,,
$$
from the table XVIII of Ref.~\cite{buras3} calculated numerically from
perturbative QCD at $\mu = 1$ GeV for $m_t = 170$ GeV in leading (LO)
and next-to-leading orders in ``naive dimensional regularization'' (NDR) 
and $^{'}$t-Hooft-Veltman (HV) regularization schemes.
   The numerical values of the QCD scale $\Lambda ^{(4)}_{\overline{MS}}$
given in table 3 correspond to 
$\alpha ^{(4)}_{\overline{MS}}(M_Z)=0.119 \pm 0.003$.
   $\xi_i^{(z)}$ and $\xi_i^{(y)}$ were obtained from $z_i$
and $y_i$, respectively, using the Eqs.~(\ref{ci-param}) and
(\ref{xi-param}).

   As we cannot calculate the factors $\widetilde{B}_i(\mu)$ theoretically,
they can be fixed only from data in the spirit of the semi-phenomenological
approach \cite{buras3,buras1,buras4}.
   Table 2 shows that the amplitudes of $K\to 2\pi$ decays
are dominated by the contribution of the operators ${\cal O}_i$ with
$i =1,2,3,4,5,8$.
   Moreover, in case of the operators ${\cal O}_{1,2,3}$, the first
term in the combination $(-\xi_1\! + \xi_2\! + \xi_3)$ dominates in
the effective weak meson lagrangian (\ref{weak-mes}).
   Thus, the isotopic amplitudes can be given after restriction to the 
dominating contributions of four-quark operators as
\begin{eqnarray}
{\cal A}_I &=& {\cal A}_I^{(z)}+\tau{\cal A}_I^{(y)}\,,
\nonumber \\
{\cal A}_I^{(z,y)}&=&
    \Big[-\xi_1^{(z,y)}(\mu)+\xi_2^{(z,y)}(\mu)+\xi_3^{(z,y)}(\mu)\Big]
     \widetilde{B}_1(\mu)\,{\cal A}_I^{(1)}
   +\xi_4^{(z,y)}(\mu)\widetilde{B}_4(\mu)\,{\cal A}_I^{(4)}
\nonumber \\ &&
   +\xi_5^{(z,y)}(\mu)\widetilde{B}_5(\mu)\,{\cal A}_I^{(5)}
   +\xi_8^{(z,y)}(\mu)\widetilde{B}_8(\mu)\,{\cal A}_I^{(8)}
    \Big]\,,
\label{AI-approximation}
\end{eqnarray}
and the relation (\ref{defa}) to the measurable amplitudes may be modified to
$$
A_I = \big( a_I^{(z)} +\tau a_I^{(y)}\big)\, e^{i\delta_I}\,.
$$
   At least two factors $\widetilde{B}_1$ and $\widetilde{B}_4$ can be 
estimated from the experimental values ${\cal A}_0^{exp} \approx 10.9$ and
${\cal A}_2^{exp} \approx 0.347$ (for fixed $\widetilde{B}_5$ and 
$\widetilde{B}_8$) while the other two (penguin) factors $\widetilde{B}_5$ and
$\widetilde{B}_8$ should be fixed from other data.
   The factors $\widetilde{B}_1$, $\widetilde{B}_4$, $\widetilde{B}_5$ and
$\widetilde{B}_8$ are the analogs of the bag factors $B_2^{(1/2)}$, 
$B_1^{(3/2)}$, $B_6^{(1/2)}$ and $B_8^{(3/2)}$, respectively, introduced in 
\cite{buras3}.

   The parameter $\varepsilon^{'}$ of direct $CP$-violation in
$K\to2\pi$ decays can be expressed by the formulae
$$
\varepsilon^{'} =
-{\omega\over\sqrt2}\,{\mbox{Im}\,a_0\over\mbox{Re}\,a_0}
  \big(1-\Omega\big)\,\mbox{e}^{i(\pi/2+\delta_2-\delta_0)}\,,\quad
\omega = \frac{\mbox{Re}\,a_2}{\mbox{Re}\,a_0}\,,\quad
\Omega = \frac{1}{\omega}\frac{\mbox{Im}\,a_2}{\mbox{Im}\,a_0}\,,
$$
and the ratio $\varepsilon^{'}/\varepsilon$ be estimated as
(recall that, experimentally,
$\varepsilon^{'}/\varepsilon \approx \mbox{Re}\,\varepsilon^{'}/\varepsilon$,
$\mbox{arg}\,\varepsilon \approx \mbox{arg}\,\varepsilon^{'}$)
\begin{equation}
\frac{\varepsilon^{'}}{\varepsilon} =
 \mbox{Im}\,\lambda_t\,\big(P_0-P_2),\quad
  P_I= \frac{\omega}{\sqrt{2}\varepsilon |V_{ud}||V_{us}|}\,
       \frac{a_I^{(y)}}{a_I^{(z)}}\,,
\label{P0P2}
\end{equation}
with
$\mbox{Im}\,\lambda_t = \mbox{Im}\,V^{*}_{ts}V_{td}
                    = |V_{ub}||V_{cb}| \mbox{sin} \delta
                    = \eta |V_{us}||V_{cb}|^2$
in the standard and the Wolfenstein parameterizations of the CKM matrix.

   Table 4 gives the estimates of $\varepsilon^{'}/\varepsilon$ from a
semi-phenomenological approach obtained after fixing the correction factors
$\widetilde{B}_1$ and $\widetilde{B}_4$ for isotopic amplitudes in the 
representation (\ref{AI-approximation}) by experimental ($CP$-conserving) 
data on $\mbox{Re}\,{\cal A}_{0,2}$, and setting 
$\widetilde{B}_5 = \widetilde{B}_8 =1$.
   We have used the matrix elements of the operators ${\cal O}_i$ displayed
in table 2 (for central values of phenomenological structure coefficients
$L_i$ given in table 1), and the theoretical values $\xi_i(\mu)$ from table 3.
   The values $(\varepsilon^{'}/\varepsilon )_{min}$ and  
$(\varepsilon^{'}/\varepsilon )_{max}$ correspond to the interval  
for $\mbox{Im}\,\lambda_t$ obtained from the phenomenological analysis of 
indirect $CP$ violation in $K\to 2\pi$ decay and $B^0-\overline B^0$ mixing
\cite{buras3,buras4}:
\begin{equation}
   0.86\cdot10^{-4} \le \mbox{Im}\,\lambda_t \le 1.71\cdot 10^{-4}\,.
\label{lambda}
\end{equation}

   Table 4 demonstrates the modification of the semi-phenomenological
estimates of the parameters $\widetilde{B}_1$, $\widetilde{B}_4$ and 
$(\varepsilon^{'}/\varepsilon )_{max}$ after successive inclusion of the
corrections at $O(p^4)$ and $O(p^6)$.
   Most important are the corrections at $O(p^4)$.
   The peculiarity of the results at $O(p^2)$ lies in the observation that all
estimates of $\varepsilon^{'}/\varepsilon$ lead to negative values.
   This is related to the fact that in the case corresponding to table 4a
the contribution of gluonic penguins to $\Delta I = 1/2$ transitions appears
to be suppressed, leading, after the interplay between gluonic and 
electromagnetic penguins, to the relation $P_0 < P_2$ for the two competing
terms in (\ref{P0P2}).
   Generally speaking, $\Delta I = 1/2$ transitions loose importance compared
to $\Delta I = 3/2$ when estimating $\varepsilon^{'}/\varepsilon$.
   The situation already changes after inclusion of the correction at $O(p^4)$,
due to relative enhancement of the matrix elements for the operator
${\cal O}_5$ (see table 4b).
   Taking into account the dependence of the Wilson coefficients on the 
renormalization scheme, after including the corrections at $O(p^4)$ and 
$O(p^6)$ we obtained the following upper and lower bounds for 
$\varepsilon^{'}/\varepsilon$ (see table 4c):
\begin{equation}
-3.2\cdot 10^{-4}\le\varepsilon^{'}/\varepsilon\le 3.3\cdot 10^{-4}\,,
\label{central}
\end{equation}
where the range characterizes the uncertainty from short-distance physics.

    Our calculations have shown that especially the penguin matrix
elements are most sensitive to various refinements: higher-order derivative
terms in chiral lagrangians, the reduction of meson resonances,
$\pi^0 - \eta - \eta^{'}$ mixing, and meson loop corrections.
    It should be added that the modification of penguin matrix
elements, discussed in this note, is much more important for gluonic
than for electromagnetic penguin transitions.
    This is obvious from the observation that the latter at the lowest
order contain terms of $O(p^0)$ which remain unchanged when taking
into account the additional terms derived from the effective lagrangian
at $O(p^4)$.
    
    We give some results concerning the dependence of the above 
semi-phenomenological estimates for $\varepsilon^{'}/\varepsilon$ on the 
choice of the penguin correction factors $\widetilde{B}_5$ and 
$\widetilde{B}_8$ (figure 2) and on the values of the structure constants 
$L_i$ (figure 3).
    In figure 3 we show the dependencies of $\varepsilon^{'}/\varepsilon$
on the coefficients $L_4$, $L_5$ and $L_8$ only, to demonstrate the 
appreciable sensitivity to the variation of these parameters within their
phenomenological bounds given in table 1.
    It is caused by the fact, that the coefficients $L_4$, $L_5$ and $L_8$ 
appear in penguin contributions to the $K\to 2\pi$ amplitudes already at 
$O(p^2)$ (see (\ref{A02-p2})) while all other structure coefficients given in 
table 1 appear in the amplitudes of higher orders.

   To study the upper and lower bounds for $\varepsilon^{'}/\varepsilon$ 
corresponding to the variation of the parameters $L_i$ and
$\mbox{Im}\,\lambda_t$ within their phenomenological bounds, we have used the 
so called ``scanning''  and ``Gaussian'' methods \cite{buras4}.
   In the first case the parameters $L_i$ were scanned independently 
within the intervals defined by their central values and errors given in 
table 1 while the parameter $\mbox{Im}\,\lambda_t$ was scanned within 
phenomenological bounds (\ref{lambda}).
   In the second case we calculated the probability density distribution
for $\varepsilon^{'}/\varepsilon$ obtained by using Gaussian distributions
for the parameters $L_i$ with errors given in table 1 
\footnote{ With exception of $L_4$ which is not determined experimentally
           and therefore taking uniform distribution inside ``theoretical''
           limits $-0.8\cdot 10^{-3}\le L_4 \le 0.2\cdot 10^{-3}$.}
while for the parameter $\mbox{Im}\,\lambda_t$ using the result
obtained in \cite{buras4}:
$$
  \mbox{Im}\,\lambda_t = (1.29 \pm 0.22)\cdot 10^{-4}\,.
$$
   In tables 5 and 6 the ``scanning'' and ``Gaussian'' results for 
$\varepsilon^{'}/\varepsilon$ are given for different values of 
$\widetilde{B}_5$ ($\widetilde{B}_8 = 1$).
   Figure 4 shows typical probability density distributions for
$\varepsilon^{'}/\varepsilon$ obtained in the Gaussian case.

   Our results demonstrate that even after taking into account all 
uncertainties related to both phenomenological input parameters and 
renormalization scheme dependence, it is still rather problematic to explain 
theoretically with $\widetilde{B}_5 = \widetilde{B}_8 = 1$ the value of 
the direct CP-violation parameter
$Re(\varepsilon^{'}/\varepsilon) = (23.0 \pm 6.5)\times 10^{-4}$
measured in the experiment NA31 at CERN \cite{NA31-CP}.
   The rather high level of direct CP-violation observed in this 
experiment was confirmed by recent measurements of KTeV at FNAL 
\cite{KTeV-CP}, $(28.0 \pm 4.1)\times 10^{-4}$, and NA48 at CERN 
\cite{NA48-CP}, $(18.5 \pm 7.3)\times 10^{-4}$.
   Taking into account the result of the experiment E731 at FNAL 
\cite{E731-CP}, $(7.4 \pm 5.9)\times 10^{-4}$, the world averaged value 
is estimated as
\begin{equation}
   Re(\varepsilon^{'}/\varepsilon) = (21.2 \pm 2.8)\times 10^{-4}\,.
\label{CP-averaged}
\end{equation}

   Finally, we give some results concerning the factor $\widetilde{B}_5$ 
required to describe the experimental value (\ref{CP-averaged}) within our  
semi-phenomenological approach.
   In figures 5, 6 and 7 we show the probability density distribution
for factors $\widetilde{B}_1$, $\widetilde{B}_4$ and $\widetilde{B}_5$, 
respectively, obtained by using Gaussian distributions for the parameters 
$L_i$, $\mbox{Im}\,\lambda_t$  and $\varepsilon^{'}/\varepsilon$.
   The parameters $\widetilde{B}_1$, $\widetilde{B}_4$ and $\widetilde{B}_5$
were defined from the experimental values of the isotopic $K\to 2\pi$
amplitudes ${\cal A}_0$, ${\cal A}_2$ and the ratio 
$\varepsilon^{'}/\varepsilon$ with $\widetilde{B}_8 = 1$.
   The dispersion of these parameter values in figures 5, 6 and 7 is caused
mainly by the uncertainties of $L_i$ and $\mbox{Im}\lambda_t$, while the
experimental error of $\varepsilon^{'}/\varepsilon$ is much less influence.
   Figure 7 demonstrates the necessity for a rather large factor 
$\widetilde{B}_5$.
  It should be emphasized, that for even larger values of $\widetilde{B}_5$,
the contribution of nonpenguin operators to the $\Delta I = 1/2$ amplitude 
are still dominating (see figure 8).
   In figure 9 the probability density plots show  the correlations 
between parameters $\widetilde{B}_1$, $\widetilde{B}_4$ and $\widetilde{B}_5$. 
   The correlations between pairs of parameters $\widetilde{B}_1$,
$\widetilde{B}_4$ and $\widetilde{B}_5$, $\widetilde{B}_4$ are caused by the 
isotopic symmetry breaking related with $\pi^0 - \eta -\eta^{'}$ mixing.
   The first plot in figure 9 demonstrates the strong correlation between
$\widetilde{B}_1$ and $\widetilde{B}_4$.
   Due to relatively small contributions of penguin operators to isotopic 
amplitudes of $K\to \pi\pi$ decays there are no visible correlations between 
$\widetilde{B}_1$, $\widetilde{B}_5$ and $\widetilde{B}_4$, $\widetilde{B}_5$.
   Figure 10 shows the correlations between $\widetilde{B}_5$ and 
$\widetilde{B}_8$ calculated for central values of the phenomenological
constants $L_i$ and $\mbox{Im}\,\lambda_t$ with
$\mbox{Re}(\varepsilon^{'}/\varepsilon) = 21.2\times 10^{-4}$ used as 
experimental input.
  From this figure one can see that even for $\widetilde{B}_8=0$ values of
$\widetilde{B}_5 > 2$ are necessary to explain the large value
of $\varepsilon^{'}/\varepsilon$ (\ref{CP-averaged}).

%============================================================================
\section{Conclusion}
%============================================================================

    From studying the impact of the recently confirmed large $\varepsilon^{'}$
value on the parameterization of the hadronic weak lagrangian, including step 
by step various refinements, we have shown the necessity for a rather large 
gluonic penguin contribution (the factor $\widetilde{B}_5$ is found well 
above 1, see figure 7). 
    The large $\widetilde{B}_1$ and $\widetilde{B}_5$ values may be a
hint that the long-distance contributions are still not completely 
understood.
    An analogous conclusion has been drawn in \cite{buras5}, where also
possible effects from physics beyond the Standard Model are discussed.
    From the phenomenological point of view, there is no difficulty
in taking (\ref{weak-mes}) as a bona-fide weak current-current lagrangian 
with coupling constants $\xi_i$ to be fixed experimentally.
    The problems arise when matching these parameters to Wilson coefficients
derived from perturbative QCD, which is, of course, a necessary requirement.
    It should be remarked that in our approach there is also no convincing
argument for the large correction factor $\widetilde{B}_1$ (due to the
$\Delta I = 1/2$ rule); but then we may ask, why $\widetilde{B}_1$ and
$\widetilde{B}_5$ should behave differently: as can be seen from table 2,
the relative changes of the respective matrix elements in going to higher 
powers of $p^2$ do not differ very much.

   In this context, one should note recent progress in the estimates
of the $B$-factors with a matching procedure based on higher-order 
calculations in the long-distance regime within the $1/N_c$-expansion.
   An essential enhancement of the bag factor for the gluonic penguin 
operator by the $1/N_c$ corrections at next-to-leading order in the chiral 
expansion has been observed in \cite{paschos2}, where the value 
$B_6^{(1/2)} = 1.6 \pm 0.1$ has been obtained.
   The similar value, $B_6^{(1/2)} = 1.6 \pm 0.3$, arising from $O(p^4)$ 
chiral loop corrections, was obtained in \cite{bertolini1,bertolini2} within
the semiphenomenological chiral quark model with values of the quark and gluon 
condensates fixed by reproducing the $\Delta I = 1/2$ rule.

    Since our results are very sensitive to the relative contribution
of the gluonic penguin operator, the question of its phenomenological
separation in $K\to 2\pi$ decays becomes critical, in the context
of the $\Delta I = 1/2$ rule as well as for a very important problem of
direct $CP$-violation.
    $CP$-conserving $K\to 2\pi$ data alone are clearly not sufficient
for such a separation.
    It could be accomplished, on the other hand, when taking into account
Dalitz-plot data for $K\to 3\pi$ as well as differential distributions
for radiative decays $K\to 2\pi\gamma$, $K\to \pi 2\gamma$ described by 
the same lagrangian (\ref{weak-lagr}).
    As emphasized above, the reason for this possibility is found in the 
difference in momentum power counting behavior between penguin and 
non-penguin matrix elements, which appears in higher orders of the chiral 
theory, when calculating various parameters of differential distributions, 
for instance, slope parameters of the Dalitz-plot for $K\to 3\pi$.
    A substantial improvement in the accuracy of such experimental data 
(mostly being of older dates) would be very helpful for such a 
phenomenological improvement of the theoretical situation for
$\varepsilon^{'}/\varepsilon$ (see \cite{bos-weak} for discussion
of this point and \cite{hyperon,tnf} for some recent measurements).

    The authors gratefully acknowledge fruitful and helpful discussions with
W.A.~Bar\-deen, A.J.~Buras, J.~Gasser, E.A.~Paschos and P.H.~Soldan.

%===========================================================================
%                               References
%===========================================================================

\newpage

%============================================================================
%\section*{Figure caption}
%============================================================================

%----------------------------------------------
\hspace*{-10mm}
\epsfxsize=80mm \epsfbox{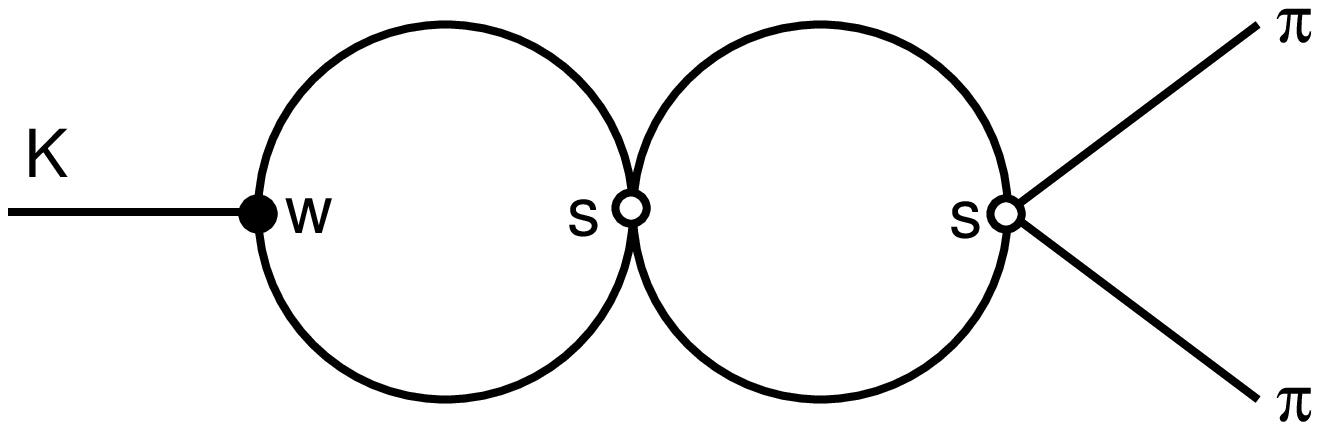}
\epsfxsize=80mm \epsfbox{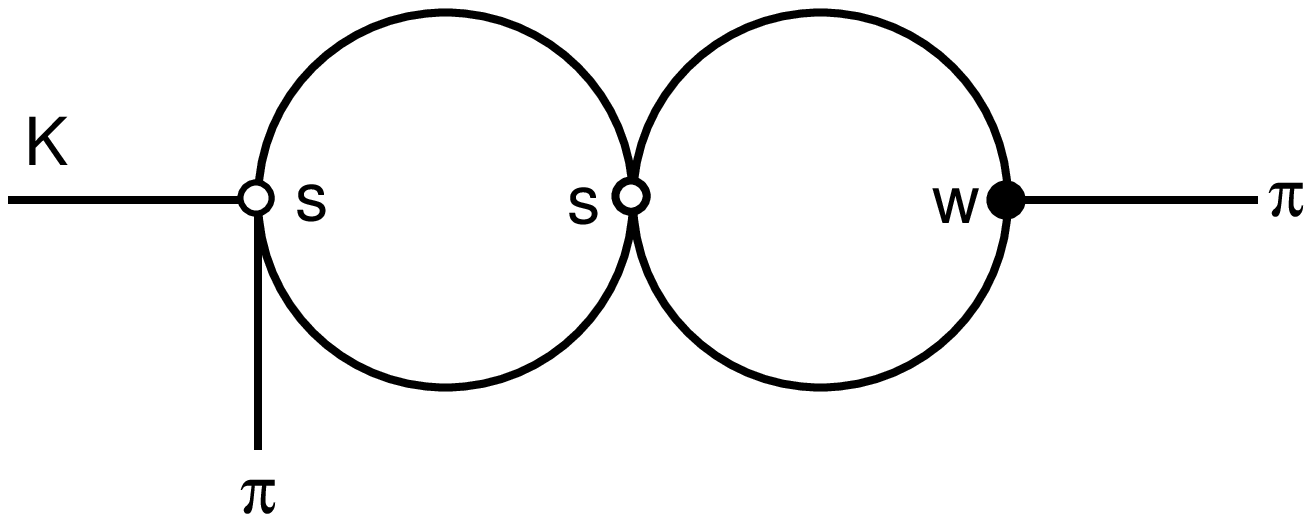}
\\
\hspace*{0.45\textwidth}{\Large\bf a)} \\
%----------------------------------------------
\epsfxsize=47mm \epsfbox{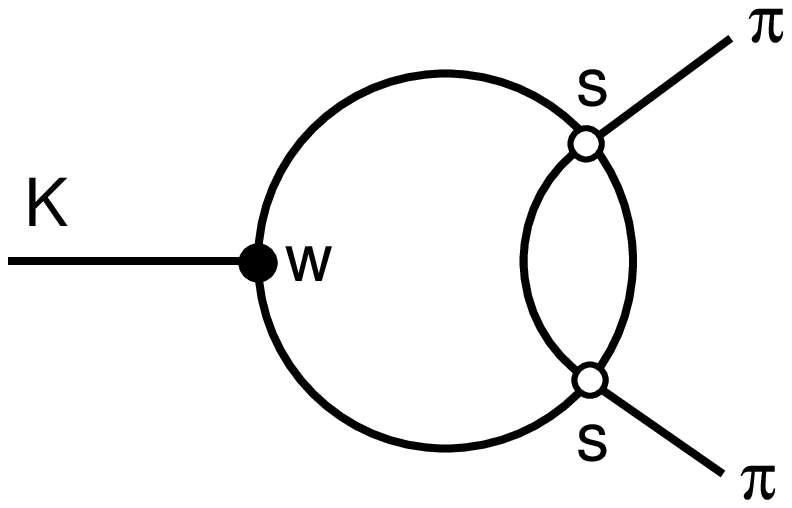}
\hspace*{33mm}\raisebox{-5mm}{\epsfxsize=80mm \epsfbox{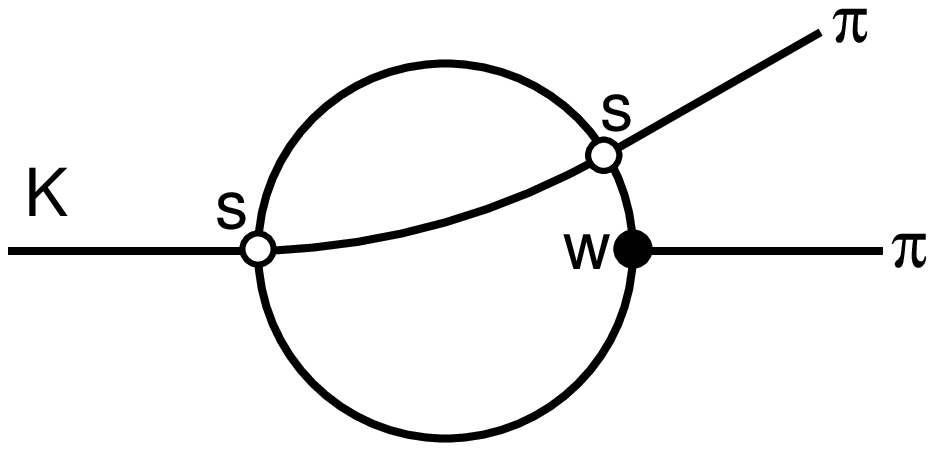}}
\\[-1ex]
\hspace*{0.45\textwidth}{\Large\bf b)} \\
%----------------------------------------------
\epsfxsize=80mm \epsfbox{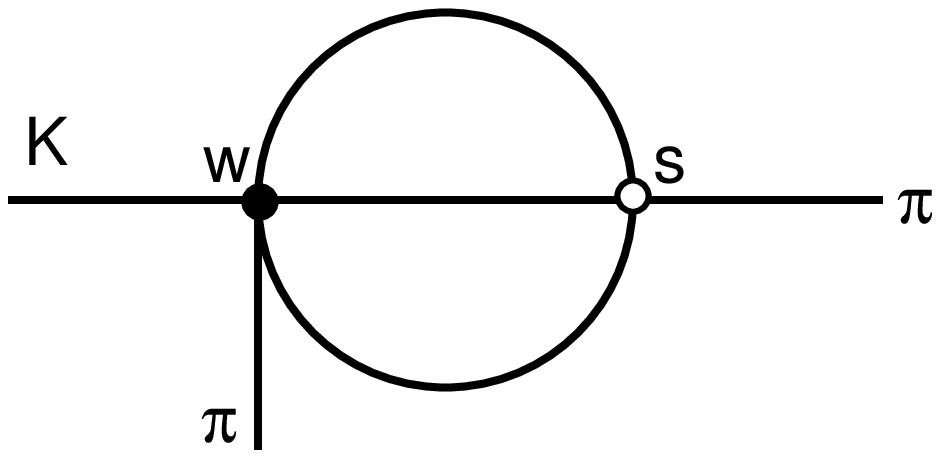}
\epsfxsize=80mm \epsfbox{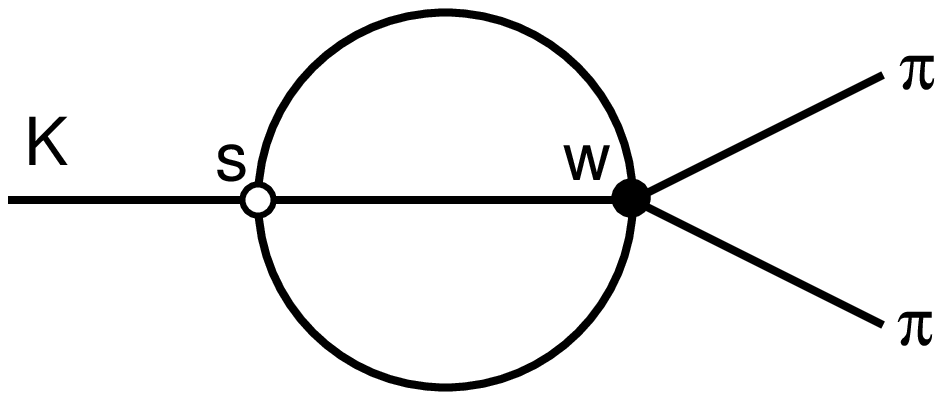}
\\[-1ex]
\hspace*{0.45\textwidth}{\Large\bf c)} \\
%----------------------------------------------
\epsfxsize=80mm \epsfbox{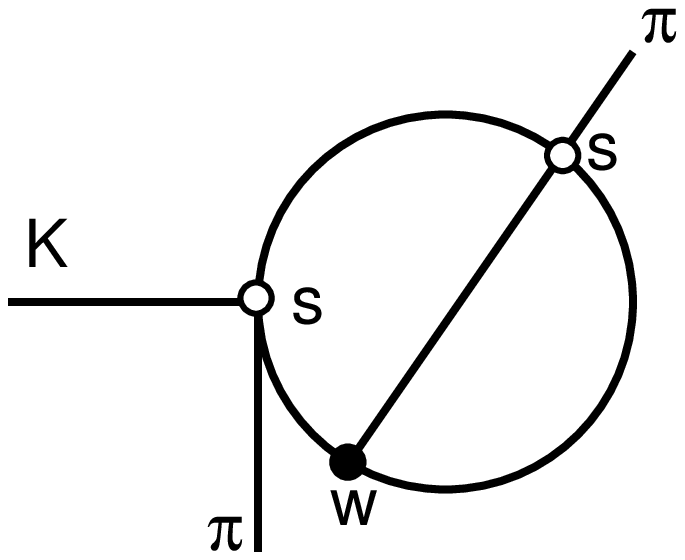}
\epsfxsize=80mm \epsfbox{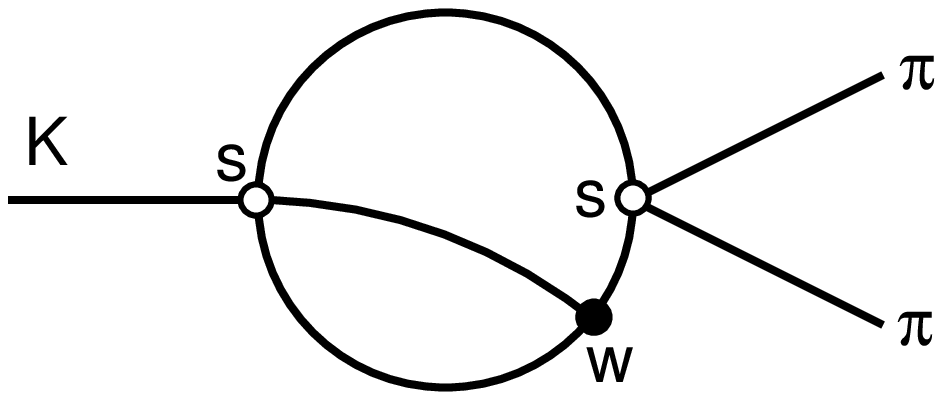}
\\[-1ex]
\hspace*{0.45\textwidth}{\Large\bf d)} \\
%----------------------------------------------
{\bf Figure 1.} Topology of the main two-loop diagrams at $O(p^6)$ (diagrams 
                with tadpole loops on the external lines and in the vertices
                are not shown). The external lines denote the momenta. The 
                internal lines correspond to various combinations of virtual 
                pions and kaons in different charge channels. The filled 
                circle denotes the week interaction vertex, the open circle 
                corresponds to strong interaction.
\newpage
%============================================================================

\epsfysize=0.9\textheight
\epsfbox{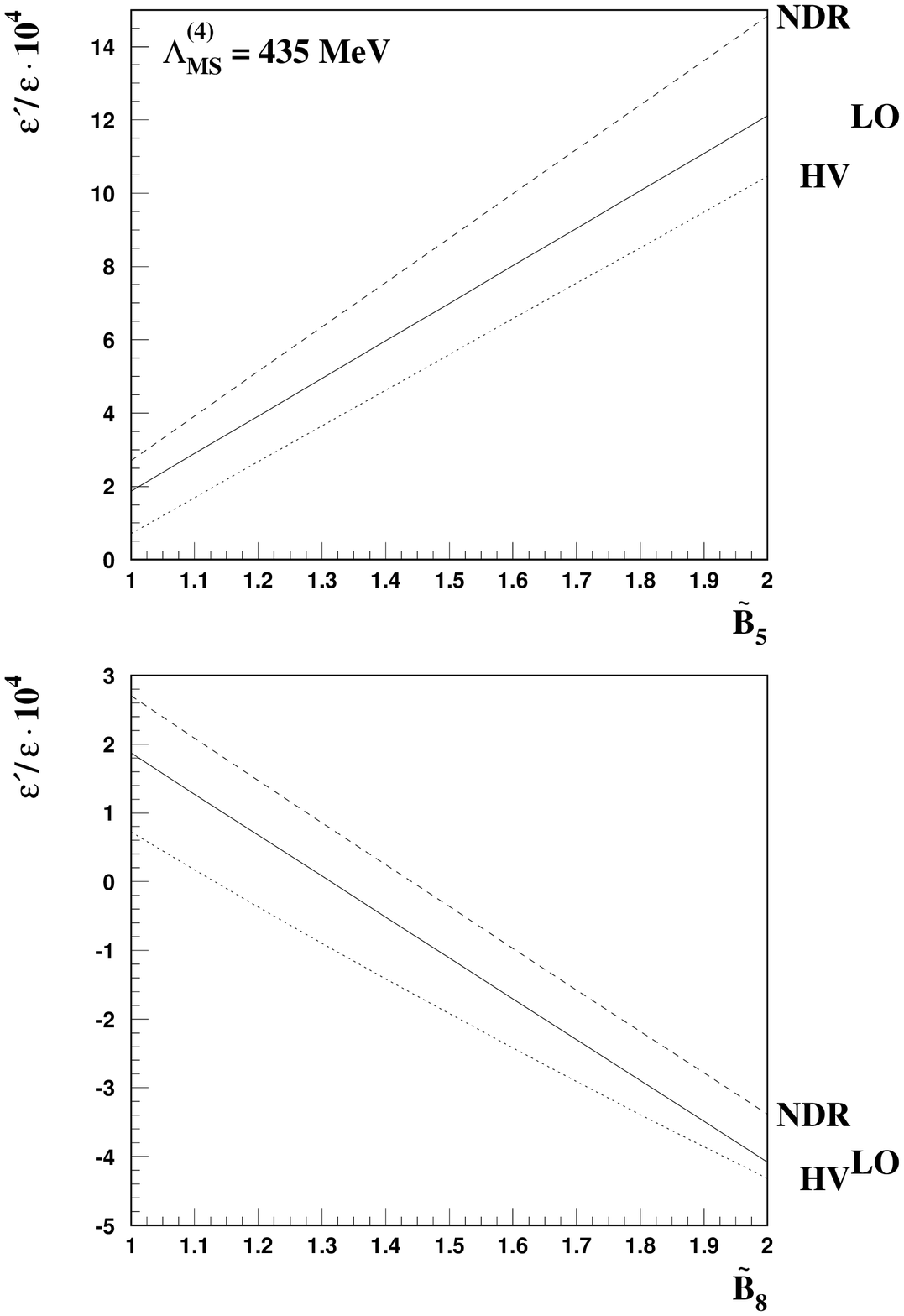}

{\bf Figure 2.}  $\widetilde{B}_5$ and $\widetilde{B}_8$-dependence of $\varepsilon^{'}/\varepsilon$
                 calculated for central values of the phenomenological 
                 constants $L_i$ and $\mbox{Im}\,\lambda_t = 1.29\cdot 10^{-4}$
                 with $\Lambda^{(4)}_{\overline{MS}}=435$ MeV. 
                 The $\widetilde{B}_5$-dependence is calculated with 
                 $\widetilde{B}_8=1$ and the $\widetilde{B}_8$-dependence -- 
                 with $\widetilde{B}_5=1$.
\newpage
%============================================================================

\epsfysize=0.9\textheight
\epsfbox{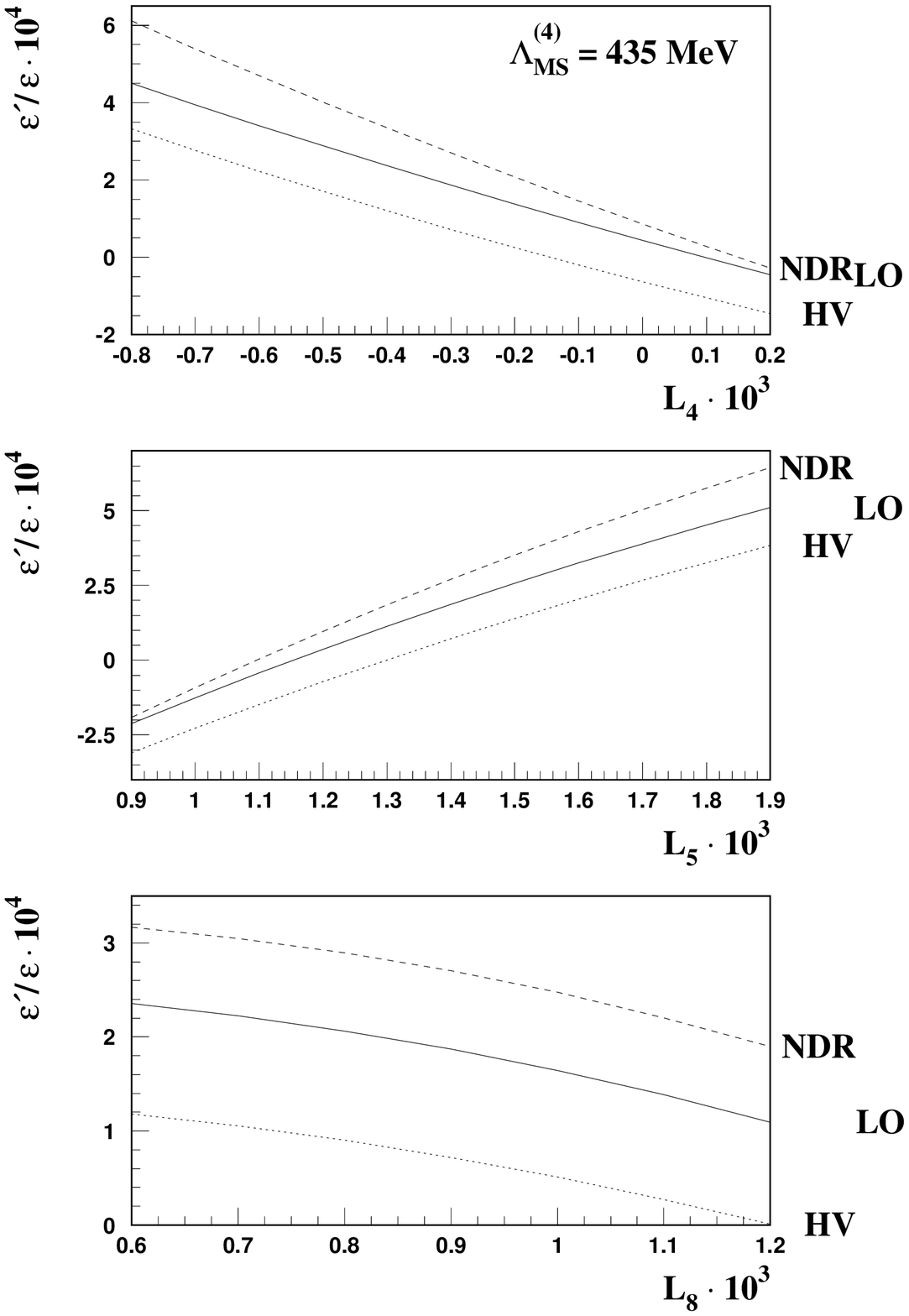}

{\bf Figure 3.}  Dependencies of $\varepsilon^{'}/\varepsilon$ on $L_4$, 
                 $L_5$, and $L_8$ calculated for central values of 
                 $\mbox{Im}~\lambda_t$ and other $L_i$-coefficients with 
                 $\Lambda^{(4)}_{\overline{MS}}=435$ MeV and 
                 $\widetilde{B}_5=\widetilde{B}_8=1$.
\newpage
%============================================================================

\epsfysize=0.9\textheight
\epsfbox{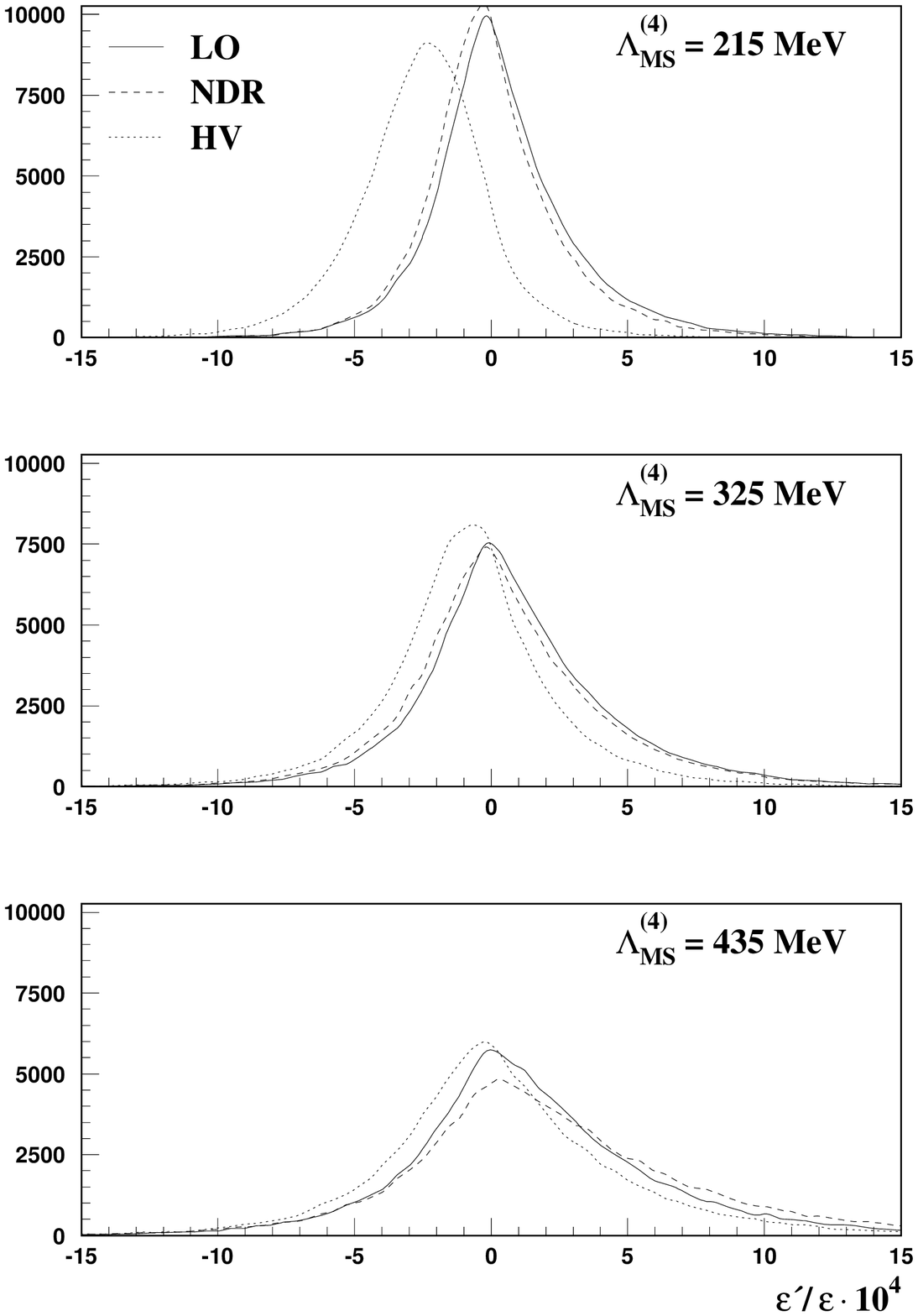}

{\bf Figure 4.}  Probability density distributions for 
                 $\varepsilon^{'}/\varepsilon$ with 
                 $\widetilde{B}_5=\widetilde{B}_8=1$. 
\newpage
%============================================================================

\epsfysize=0.9\textheight
\epsfbox{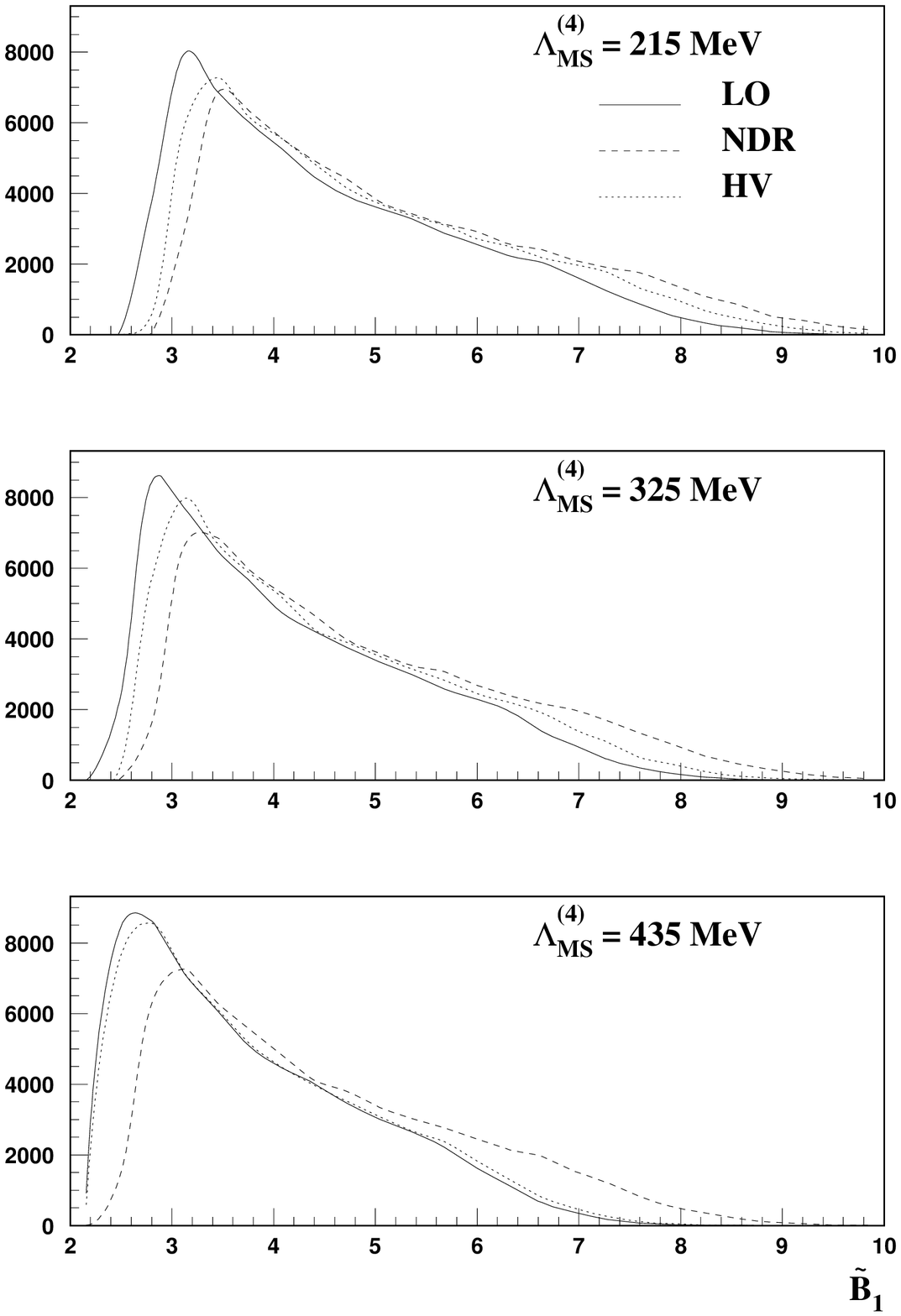}

{\bf Figure 5.}  Probability density distributions for 
                 $\widetilde{B}_1$ with $\widetilde{B}_8=1$. 
\newpage
%============================================================================

\epsfysize=0.9\textheight
\epsfbox{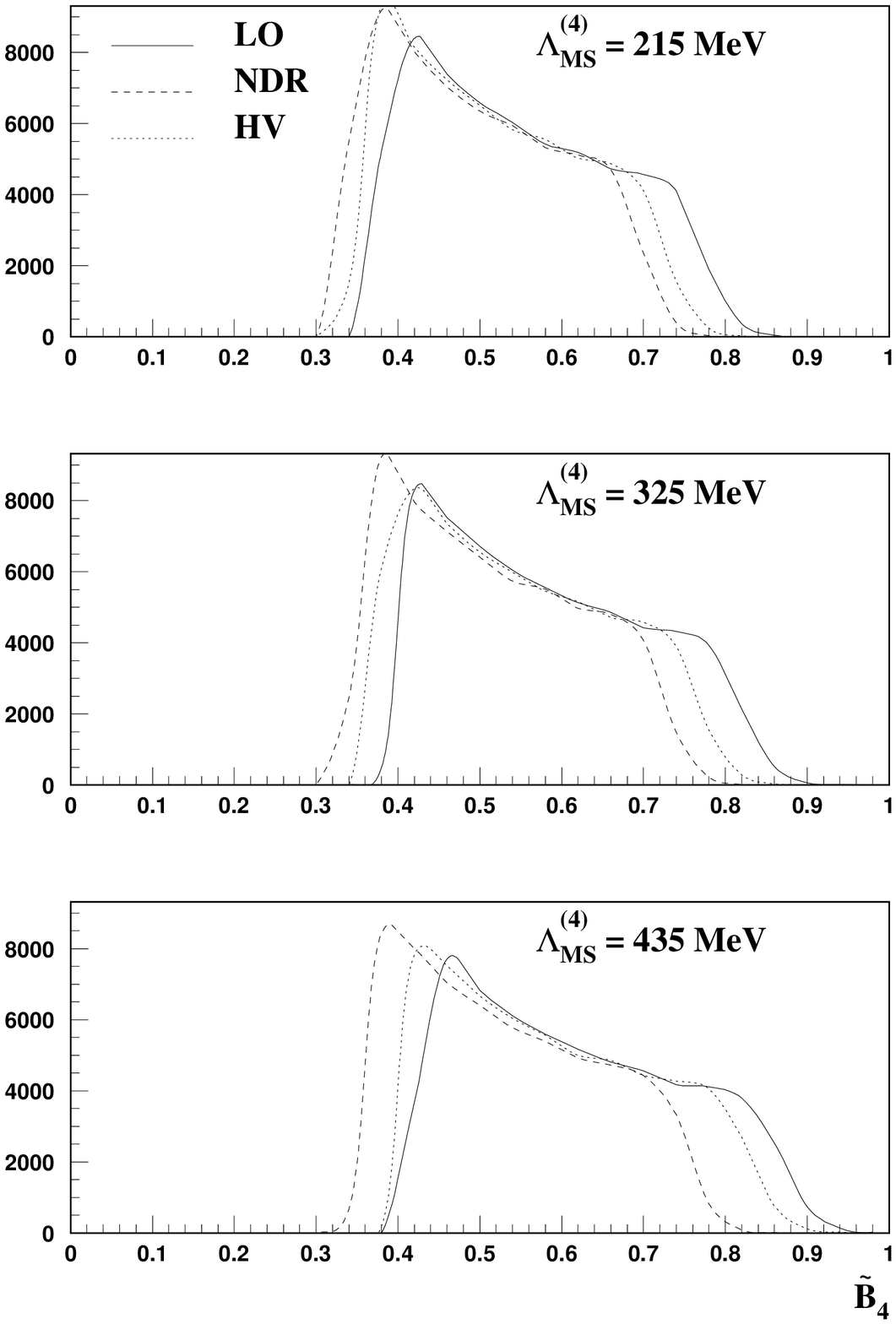}

{\bf Figure 6.}  Probability density distributions for 
                 $\widetilde{B}_4$ with $\widetilde{B}_8=1$. 
\newpage
%============================================================================

\epsfysize=0.9\textheight
\epsfbox{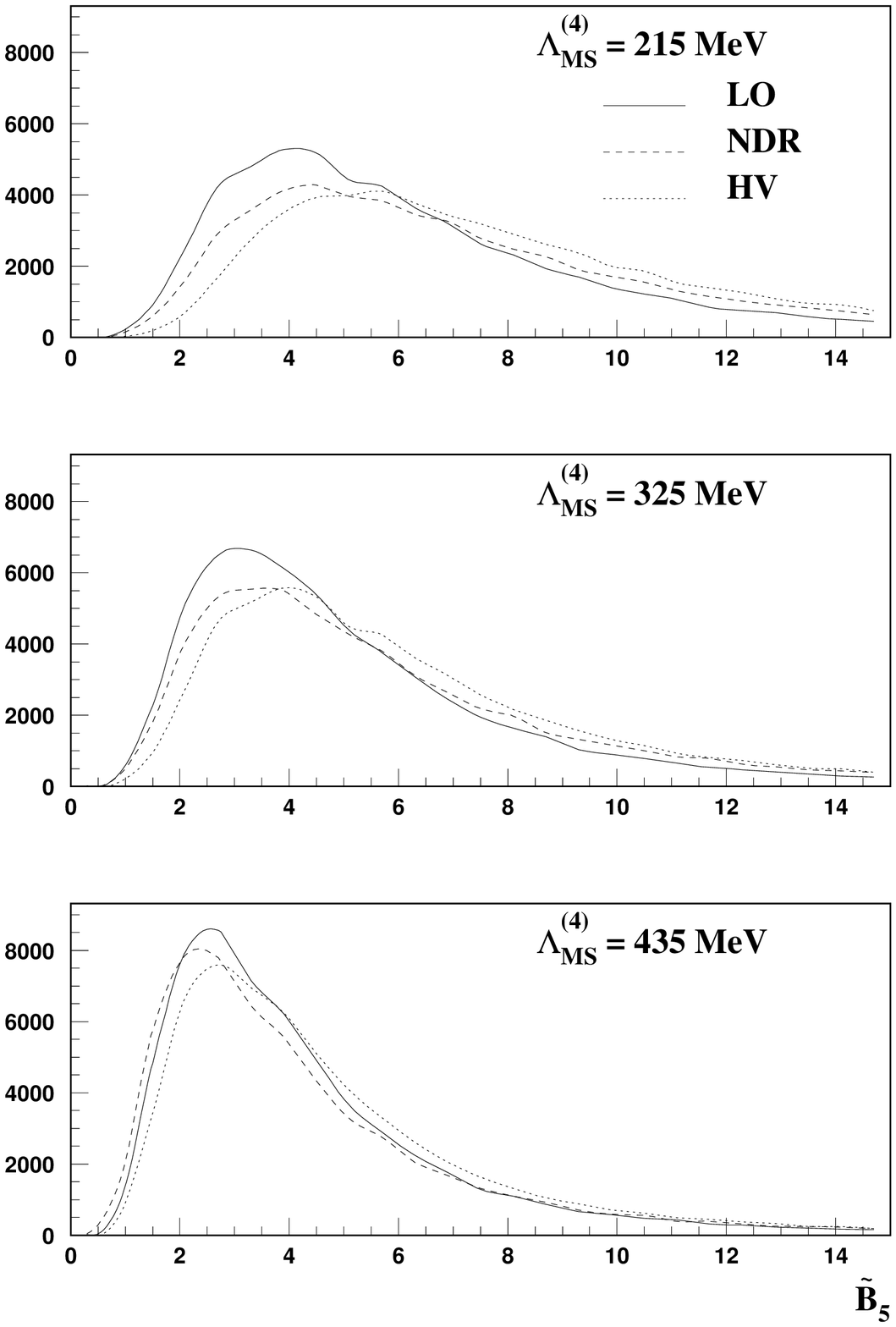}

{\bf Figure 7.}  Probability density distributions for 
                 $\widetilde{B}_5$ with $\widetilde{B}_8=1$. 
\newpage
%============================================================================

\epsfxsize=0.9\textwidth
\epsfbox{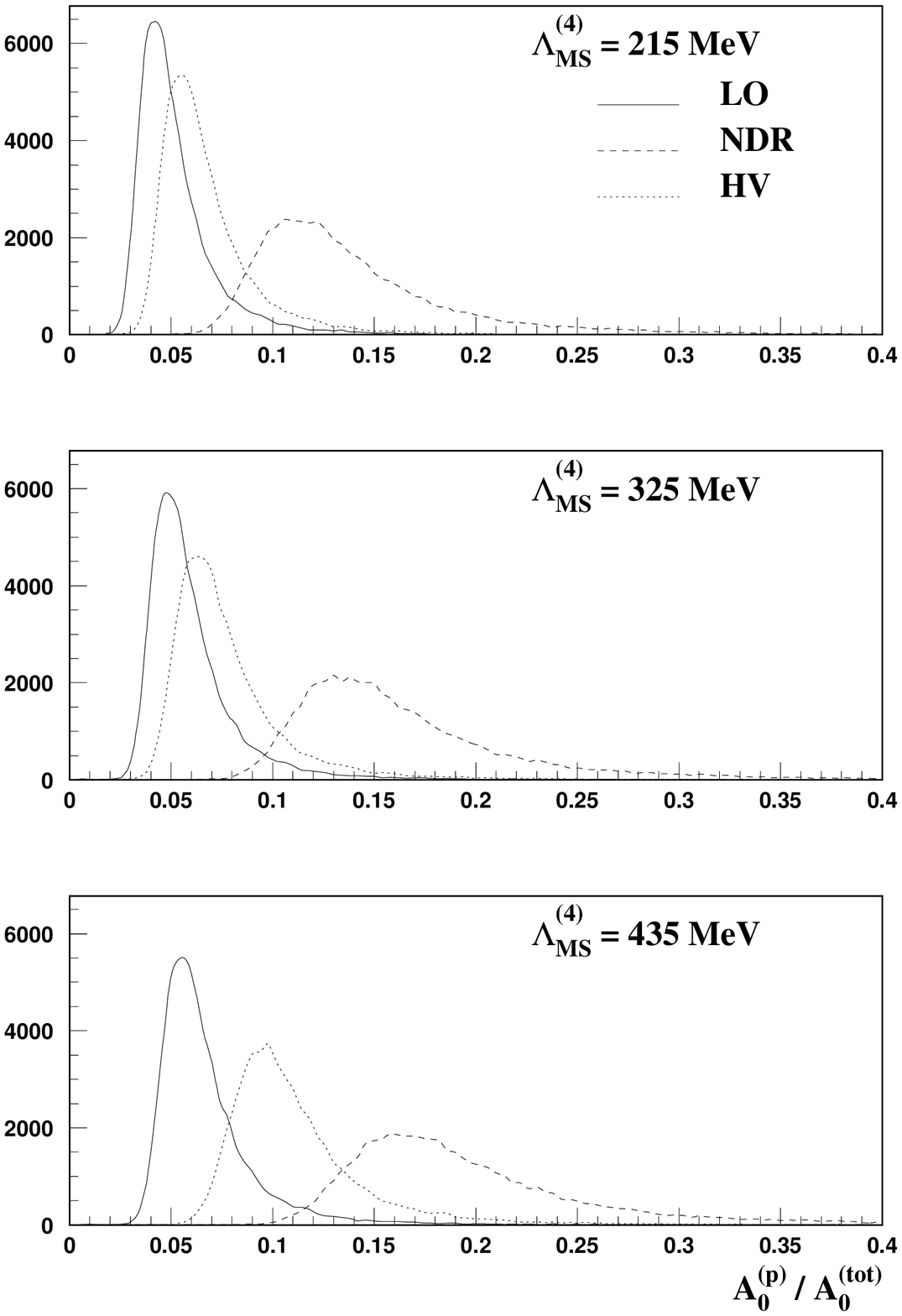}

{\bf Figure 8.} Probability density distributions for the relative 
                contribution of penguin operators to the $\Delta I =1/2$ 
                amplitude.
\newpage
%============================================================================

\epsfxsize=\textwidth
\epsfbox{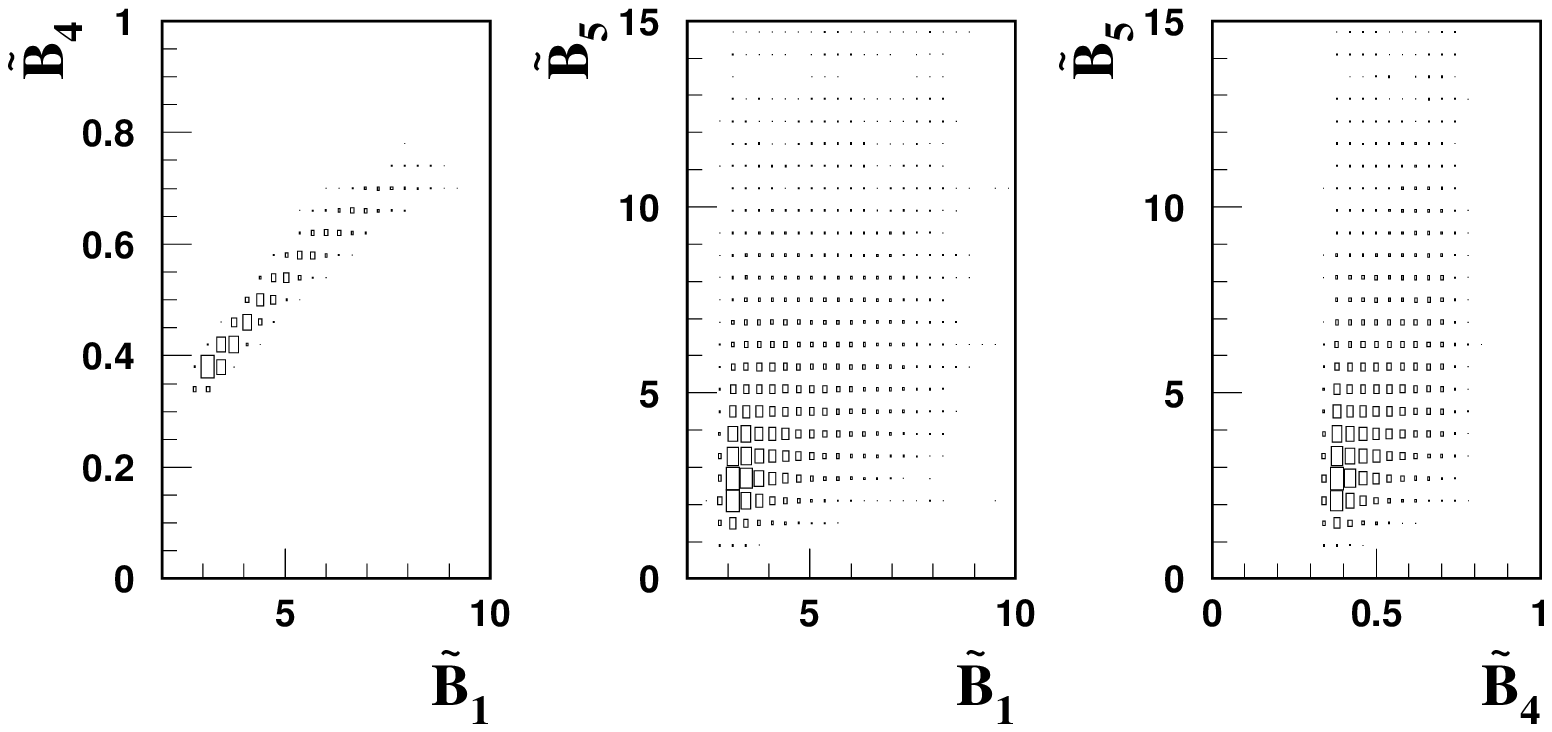}

{\bf Figure 9.}  Typical probability density plots for correlations between 
                 parameters $\widetilde{B}_1$, $\widetilde{B}_4$ and 
                 $\widetilde{B}_5$ calculated with $\widetilde{B}_8=1$ 
                 and $\Lambda ^{(4)}_{\overline{MS}}=325$ MeV in NDR scheme. 
                 For various values of $\Lambda ^{(4)}_{\overline{MS}}$ and
                 different renormalization schemes such plots look very 
                 similar.
\newpage
%============================================================================

\epsfxsize=\textwidth
\epsfbox{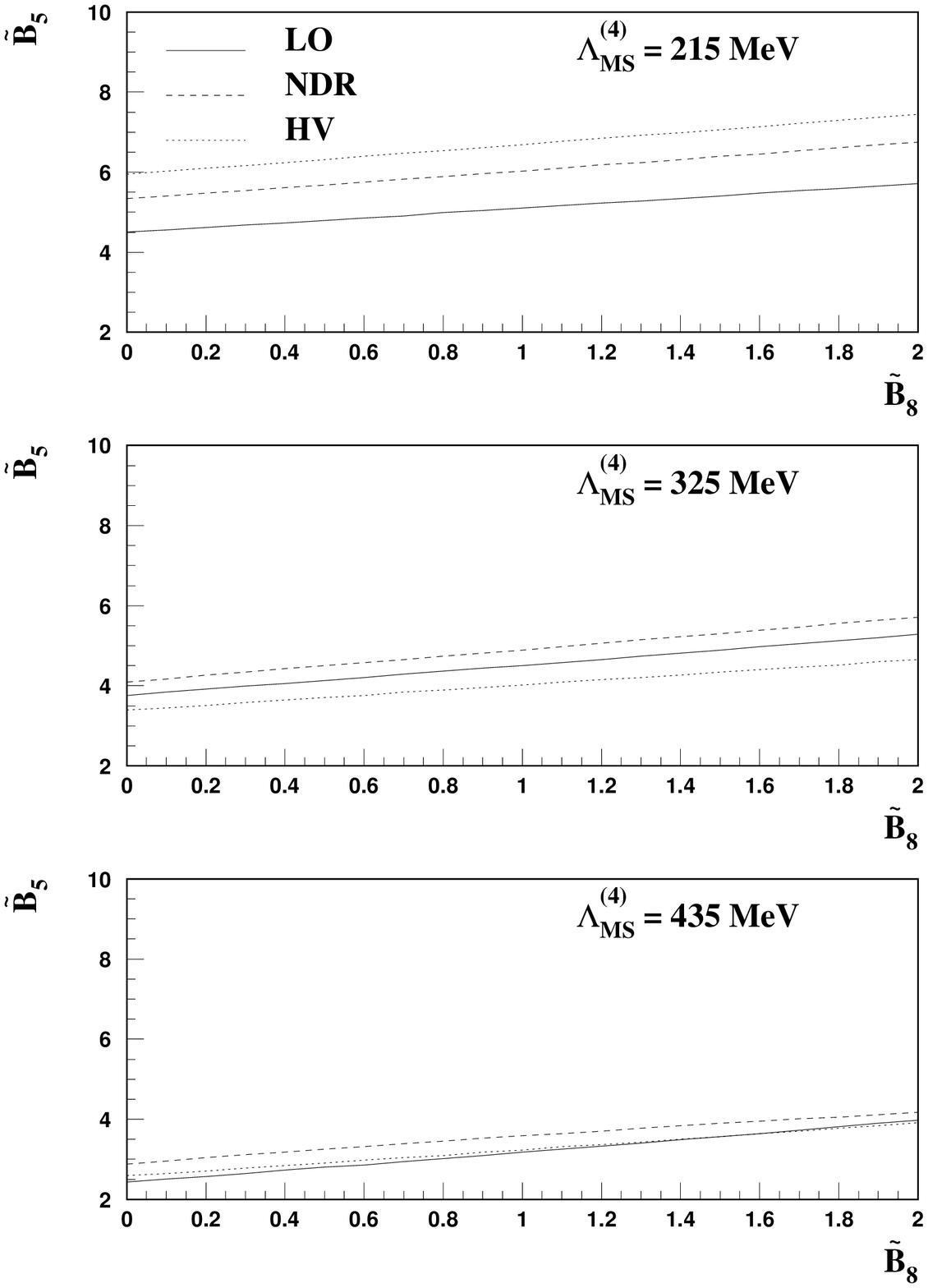}

{\bf Figure 10.}  Correlations between parameters $\widetilde{B}_5$ and 
                  $\widetilde{B}_8$.
\newpage
%=============================================================================
%                            Table 1
%=============================================================================
{\bf Table 1.} Phenomenological and theoretical values of the structure 
               coefficients $L_i$  (in units $10^{-3}$).
\\

%=============================================================================
\begin{center}
\small{
\begin{tabular}{|c|c|c|c|c|c|}
\hline
$L_i$& \multicolumn{3}{|c|}{Phenomenology \cite{dafne}}
     &\multicolumn{2}{|c|}{NJL model} \\ \cline{2-6}
     &$L_i^r(m_\rho)$&Input&$\Gamma_i$&Without reduction&After reduction\\[-1ex]
     &               &     &          &  of resonances  &  of resonances\\
\hline
1& $ 0.4 \pm 0.3 $& $K_{e4}$ and $\pi\pi\to\pi\pi$& 3/32 & 0.79 & 0.85 \\
2& $ 1.4 \pm 0.3 $& $K_{e4}$ and $\pi\pi\to\pi\pi$& 3/16 & 1.58 & 1.70 \\
3& $-3.5 \pm 1.1 $& $K_{e4}$ and $\pi\pi\to\pi\pi$&   0  &-3.17 &-4.30 \\
4& $-0.3 \pm 0.5 $& $1/N_c$ arguments             & 1/8  &  0   &  0   \\
5& $ 1.4 \pm 0.5 $& $F_K / F_\pi$                 & 3/8  & 0.98 & 1.64 \\
8& $ 0.9 \pm 0.3 $& $m_{K^0}-m_{K^+}$, $L_5$,     & 5/48 & 0.36 & 1.12 \\[-1ex]
 &                &  baryon mass ratios           &      &      &      \\
\hline 
\end{tabular}
}
\end{center}
%-----------------------------------------------------------------------------
\newpage

%=============================================================================
%                                 Table 2
%=============================================================================
\vspace*{-16mm}
\begin{center}
{\bf Table 2.} Isotopic amplitudes of $K\to2\pi$ decays
\end{center}

\vspace*{-1ex}
{\small
\begin{tabular}{|l|l|*{8}{r}|} \hline
%\multicolumn{2}{|c}{}
Operators&\multicolumn{1}{c}{}
                   &${\cal O}_1$ &${\cal O}_2$ &${\cal O}_3$ &${\cal O}_4$
                   &${\cal O}_5$ &${\cal O}_6$
                   &$\alpha {\cal O}_7$ &$\alpha {\cal O}_8$\\
\hline
%===========================================================================p^2
&\multicolumn{9}{|c|}{Soft pion approximation}\\ \cline{2-10}
&$\mbox{Re} {\cal A}_0^{(i)}$
         &\hspace*{-2mm}--1.000 & 1.000 & 1.000 & 0.000 &\hspace*{-2mm}--9.623 & 0.000 & 0.016 & 1.458\\
$~~O(p^2)$&$\mbox{Re} {\cal A}_2^{(i)}$
         & 0.000 & 0.000 & 0.000 & 1.000 & 0.000 & 0.000 &\hspace*{-2mm}--0.016 & 0.654\\[-0.35ex]
\cline{2-10}
&\multicolumn{9}{|c|}{Born diagrams with $\pi_0-\eta-\eta^\prime$ mixing}\\ 
\cline{2-10}
&$\mbox{Re} {\cal A}_0^{(i)}$
         & 0.004 &\hspace*{-2mm}\hspace*{-2mm}--0.021 &\hspace*{-2mm}--0.039 & 0.020 & 0.119 & 0.004 &\hspace*{-2mm}--0.001 &\hspace*{-2mm}--0.002\\[-0.35ex]
&$\mbox{Re} {\cal A}_2^{(i)}$
         &\hspace*{-2mm}--0.004 & 0.021 & 0.039 &\hspace*{-2mm}--0.002 &\hspace*{-2mm}--0.119 &\hspace*{-2mm}--0.004 &0.000 &\hspace*{-2mm}--0.016\\
\hline
\hline
~~~~Sum&$\mbox{Re} {\cal A}_0^{(i)}$
         &\hspace*{-2mm}--0.996 & 0.979 & 0.961 & 0.020 &\hspace*{-2mm}--9.504 & 0.004 & 0.015 & 1.456\\[-0.35ex]
$~~~~~~p^2$&$\mbox{Re} {\cal A}_2^{(i)}$
         &\hspace*{-2mm}--0.004 & 0.021 & 0.039 & 0.998 &\hspace*{-2mm}--0.099 &\hspace*{-2mm}--0.004 &\hspace*{-2mm}--0.016 & 0.638\\
\hline
\hline
%===========================================================================p^4
&\multicolumn{9}{|c|}{Born diagrams }\\  \cline{2-10}
&$\mbox{Re} {\cal A}_0^{(i)}$
         &\hspace*{-2mm}--0.247 & 0.249 & 0.236 & 0.008 &\hspace*{-2mm}--1.626 & 0.000 & 0.004 & 0.037\\[-0.35ex]
&$\mbox{Re} {\cal A}_2^{(i)}$
         &\hspace*{-2mm}--0.003 & 0.001 & 0.015 & 0.249 &\hspace*{-2mm}--0.059 & 0.000 &\hspace*{-2mm}--0.004 & 0.008\\
\cline{2-10}
$~~O(p^4)$&\multicolumn{9}{|c|}{1-loop diagrams }\\  \cline{2-10}
&$\mbox{Re} {\cal A}_0^{(i)}$
         &\hspace*{-2mm}--0.171 & 0.171 & 0.111 & 0.001 &\hspace*{-2mm}--2.072 & 0.000 & 0.001 & 0.188\\[-0.35ex]
&$\mbox{Im} {\cal A}_0^{(i)}$
         &\hspace*{-2mm}--0.482 & 0.482 & 0.482 & 0.000 &\hspace*{-2mm}--4.572 & 0.000 & 0.008 & 0.344\\[-0.35ex]
&$\mbox{Re} {\cal A}_2^{(i)}$
         & 0.000 & 0.000 &\hspace*{-2mm}--0.004 &\hspace*{-2mm}--0.149 & 0.001 & 0.000 & 0.001 &\hspace*{-2mm}--0.006\\[-0.35ex]
&$\mbox{Im} {\cal A}_2^{(i)}$
         & 0.000 & 0.000 & 0.000 &\hspace*{-2mm}--0.213 &\hspace*{-2mm}--0.004 & 0.000 & 0.003 &\hspace*{-2mm}--0.049\\
\hline
\hline
&$\mbox{Re} {\cal A}_0^{(i)}$
         &\hspace*{-2mm}--1.415 & 1.399 & 1.307 & 0.029 &\hspace*{-2mm}--13.202& 0.004 & 0.020 & 1.871\\[-0.35ex]
~~~~Sum&$\mbox{Im} {\cal A}_0^{(i)}$
         &\hspace*{-2mm}--0.482 & 0.482 & 0.482 & 0.000 &\hspace*{-2mm}--4.572 & 0.000 & 0.008 & 0.806\\[-0.35ex]
$~~~p^2+p^4~$&$\mbox{Re} {\cal A}_2^{(i)}$
         &\hspace*{-2mm}--0.007 & 0.022 & 0.050 & 1.099 &\hspace*{-2mm}--0.157 &\hspace*{-2mm}--0.004 &\hspace*{-2mm}--0.018 & 0.593\\[-0.35ex]
&$\mbox{Im} {\cal A}_2^{(i)}$
         & 0.000 & 0.000 & 0.000 &\hspace*{-2mm}--0.213 &\hspace*{-2mm}--0.004 & 0.000 & 0.003 &\hspace*{-2mm}--0.151\\
\hline
\hline
%===========================================================================p^6
&\multicolumn{9}{|c|}{Born diagrams }\\  \cline{2-10}
&$\mbox{Re} {\cal A}_0^{(i)}$
         &\hspace*{-2mm}--0.003 & 0.005 & 0.005 & 0.000 & 0.012 & 0.000 & 0.000 & 0.001\\[-0.35ex]
&$\mbox{Re} {\cal A}_2^{(i)}$
         &\hspace*{-2mm}--0.001 &\hspace*{-2mm}\hspace*{-2mm}--0.001 &\hspace*{-2mm}--0.001 & 0.005 &\hspace*{-2mm}--0.004 & 0.000 & 0.000 & 0.000\\
\cline{2-10}
&\multicolumn{9}{|c|}{1-loop diagrams }\\  \cline{2-10}
&$\mbox{Re} {\cal A}_0^{(i)}$
         &\hspace*{-2mm}--0.106 & 0.107 & 0.018 & 0.002 &\hspace*{-2mm}--0.151 & 0.000 &\hspace*{-2mm}--0.002 & 0.016\\[-0.35ex]
$~~O(p^6)$&$\mbox{Im} {\cal A}_0^{(i)}$
         &\hspace*{-2mm}--0.229 & 0.232 & 0.232 & 0.000 &\hspace*{-2mm}--1.582 &\hspace*{-2mm}--0.001 & 0.004 & 0.063\\[-0.35ex]
&$\mbox{Re} {\cal A}_2^{(i)}$
         & 0.000 & 0.001 & 0.002 &\hspace*{-2mm}--0.097 &\hspace*{-2mm}--0.004 & 0.000 & 0.000 & 0.007\\[-0.35ex]
&$\mbox{Im} {\cal A}_2^{(i)}$
         & 0.000 & 0.001 & 0.001 &\hspace*{-2mm}--0.077 &\hspace*{-2mm}--0.001 & 0.000 & 0.001 &\hspace*{-2mm}--0.007\\
\cline{2-10}
&\multicolumn{9}{|c|}{2-loop diagrams }\\  \cline{2-10}
&$\mbox{Re} {\cal A}_0^{(i)}$
         & 0.202 &\hspace*{-2mm}\hspace*{-2mm}--0.202 &\hspace*{-2mm}--0.220 & 0.000 & 1.753 & 0.000 &\hspace*{-2mm}--0.003 &\hspace*{-2mm}--0.075\\[-0.35ex]
&$\mbox{Im} {\cal A}_0^{(i)}$
         &\hspace*{-2mm}--0.169 & 0.169 & 0.142 & 0.000 &\hspace*{-2mm}--1.704 & 0.000 & 0.003 & 0.115\\[-0.35ex]
&$\mbox{Re} {\cal A}_2^{(i)}$
         & 0.001 &\hspace*{-2mm}\hspace*{-2mm}--0.001 &\hspace*{-2mm}--0.001 &\hspace*{-2mm}--0.036 & 0.000 & 0.000 & 0.001 &\hspace*{-2mm}--0.007\\[-0.35ex]
&$\mbox{Im} {\cal A}_2^{(i)}$
         & 0.000 & 0.000 & 0.000 & 0.034 & 0.000 & 0.000 &\hspace*{-2mm}--0.001 & 0.006\\
\hline
\hline
&$\mbox{Re} {\cal A}_0^{(i)}$
         &\hspace*{-2mm}--1.322 & 1.309 & 1.111 & 0.031 &\hspace*{-2mm}--11.588 & 0.003 & 0.015 & 1.664\\[-0.35ex]
~~~~Sum&$\mbox{Im} {\cal A}_0^{(i)}$
         &\hspace*{-2mm}--0.880 & 0.883 & 0.856 & 0.000 &\hspace*{-2mm}--7.858 &\hspace*{-2mm}--0.001 & 0.014 & 1.181\\[-0.35ex]
$p^2+p^4+p^6$&$\mbox{Re} {\cal A}_2^{(i)}$
         &\hspace*{-2mm}--0.007 & 0.021 & 0.049 & 0.971 &\hspace*{-2mm}--0.166 &\hspace*{-2mm}--0.004 &\hspace*{-2mm}--0.018 & 0.566\\[-0.35ex]
&$\mbox{Im} {\cal A}_2^{(i)}$
         & 0.000 & 0.001 & 0.001 &\hspace*{-2mm}--0.256 &\hspace*{-2mm}--0.003 & 0.000 & 0.004 &\hspace*{-2mm}--0.140\\
\hline
\end{tabular}
}
\newpage

%=============================================================================
%                                 Table 3
%=============================================================================
{\bf Table 3.} QCD predictions for the parameters
               $\xi_i(\mu)=\xi^{(z)}_i(\mu)+\tau\xi^{(y)}_i(\mu)$, calculated 
               with Wilson coefficients $c_i(\mu)=z_i(\mu)+\tau y_i(\mu)$
               at $\mu = 1$ GeV for $m_t = 170$ GeV \cite{buras3}.

\vspace{5mm}
\small{
\begin{tabular}{|l|r|r|r||r|r|r||r|r|r|}
\hline
\multicolumn{1}{|c|}{} &
\multicolumn{3}{|c||}{$\Lambda ^{(4)}_{\overline{MS}} = 215$ MeV \rule[-3mm]{0mm}{8mm}} &
\multicolumn{3}{|c||}{$\Lambda ^{(4)}_{\overline{MS}} = 325$ MeV} &
\multicolumn{3}{|c|}{$\Lambda ^{(4)}_{\overline{MS}} = 435$ MeV}\\ \hline
\multicolumn{1}{|c|}{$~~~~~~~~$} &
\multicolumn{1}{|c|}{~{\bf LO}~  }  &
\multicolumn{1}{|c|}{~{\bf NDR}} &
\multicolumn{1}{|c||}{~{\bf HV}~}  &
\multicolumn{1}{|c|}{~{\bf LO}~}  &
\multicolumn{1}{|c|}{~{\bf NDR}} &
\multicolumn{1}{|c||}{~{\bf HV}~}  &
\multicolumn{1}{|c|}{~{\bf LO}~}  &
\multicolumn{1}{|c|}{~{\bf NDR}} &
\multicolumn{1}{|c|}{~{\bf HV}~}\\ \hline
$\xi_1^{(z)}$       &--1.286 &--1.061 &--1.165 &--1.443 &--1.159 &
                     --1.325 &--1.624 &--1.270 &--1.562 \\
$\xi_2^{(z)}$       &  0.187 &  0.195 &  0.198 &  0.172 &  0.176 &
                       0.182 &  0.157 &  0.150 &  0.165 \\
$\xi_3^{(z)}$       &  0.129 &  0.143 &  0.137 &  0.122 &  0.137 &
                       0.130 &  0.115 &  0.131 &  0.121 \\
$\xi_4^{(z)}$       &  0.645 &  0.714 &  0.687 &  0.609 &  0.684 &
                       0.650 &  0.573 &  0.654 &  0.599 \\
$\xi_5^{(z)}$       &--0.008 &--0.020 &--0.008 &--0.012 &--0.032 &
                     --0.013 &--0.016 &--0.056 &--0.023 \\
$\xi_6^{(z)}$       &  0.000 &--0.003 &  0.000 &--0.001 &--0.007 &
                     --0.001 &--0.002 &--0.021 &--0.007 \\
$\xi_7^{(z)}/\alpha$&  0.002 &  0.003 &--0.001 &  0.004 &  0.008 &
                       0.001 &  0.006 &  0.015 &  0.032 \\
$\xi_8^{(z)}/\alpha$&  0.000 &  0.002 &  0.001 &  0.001 &  0.004 &
                       0.002 &  0.001 &  0.009 &  0.067 \\
\hline
\hline
$\xi_1^{(y)}$       &  0.044 &  0.038 &  0.048 &  0.054 &  0.048 &
                       0.053 &  0.065 &  0.060 &  0.069 \\
$\xi_2^{(y)}$       &--0.028 &--0.029 &--0.030 &--0.029 &--0.033 &
                     --0.030 &--0.030 &--0.033 &--0.030 \\
$\xi_3^{(y)}$       &--0.002 &--0.002 &  0.001 &--0.002 &--0.002 &
                     --0.002 &--0.002 &--0.002 &--0.002 \\
$\xi_4^{(y)}$       &--0.009 &--0.010 &  0.004 &--0.008 &--0.009 &
                     --0.009 &--0.008 &--0.009 &--0.008 \\
$\xi_5^{(y)}$       &--0.081 &--0.076 &--0.067 &--0.109 &--0.111 &
                     --0.092 &--0.143 &--0.173 &--0.132 \\
$\xi_6^{(y)}$       &--0.033 &--0.042 &--0.021 &--0.049 &--0.076 &
                     --0.033 &--0.071 &--0.139 &--0.051 \\
$\xi_7^{(y)}/\alpha$&  0.033 &  0.004 &  0.006 &  0.044 &  0.013 &
                       0.016 &  0.057 &  0.027 &  0.032 \\
$\xi_8^{(y)}/\alpha$&  0.031 &  0.028 &  0.031 &  0.043 &  0.041 &
                       0.045 &  0.058 &  0.061 &  0.067 \\
\hline
\end{tabular}
}
\newpage

%=============================================================================
%                                 Table 4
%=============================================================================
{\bf Table 4.} Predictions for the parameters of $K\to 2\pi$ decays in the 
               semi-phenomenological approach 
               ($\widetilde{B}_5 = \widetilde{B}_8 =1$).
               The ratio $\varepsilon^{'}/\varepsilon$ is given in units 
               $10^{-4}$.

\vspace{5mm}
%=============================================================================
{\large ~~~a) At $O(p^2)$:} \\[2mm]
%
%=============================================================================
%
\small
\begin{tabular}{|c|c|c|c||c|c|c||c|c|c|}
\hline
\multicolumn{1}{|c|}{} &
\multicolumn{3}{|c||}{$\Lambda ^{(4)}_{\overline{MS}} = 215$ MeV} &
\multicolumn{3}{|c||}{$\Lambda ^{(4)}_{\overline{MS}} = 325$ MeV} &
\multicolumn{3}{|c|}{$\Lambda ^{(4)}_{\overline{MS}} = 435$ MeV\rule[-3mm]{0mm}{8mm}}
 \\ \hline
\multicolumn{1}{|c|}{$~~~~~~~~$} &
\multicolumn{1}{|c|}{$~{\bf LO}~$}  &
\multicolumn{1}{|c|}{$~{\bf NDR}~$} &
\multicolumn{1}{|c||}{$~{\bf HV}~$}  &
\multicolumn{1}{|c|}{$~{\bf LO}~$}  &
\multicolumn{1}{|c|}{$~{\bf NDR}~$} &
\multicolumn{1}{|c||}{$~{\bf HV}~$}  &
\multicolumn{1}{|c|}{$~{\bf LO}~$}  &
\multicolumn{1}{|c|}{$~{\bf NDR}~$} &
\multicolumn{1}{|c|}{$~{\bf HV}~$}\\ \hline
$\widetilde{B}_{1}$
& 6.82 & 7.74 & 7.29 & 6.26 & 7.27 & 6.65 & 5.71 & 6.76 & 5.83 \\
$\widetilde{B}_{4}$
& 0.54 & 0.48 & 0.51 & 0.57 & 0.50 & 0.53 & 0.60 & 0.52 & 0.58 \\
$P_0$
& 1.76 & 1.21 & 0.59 & 3.17 & 2.73 & 2.06 & 5.02 &  5.91& 4.29 \\
$P_2$
& 2.88 & 2.49 & 3.69 & 4.41 & 4.20 & 4.33 & 6.37 & 7.03 & 6.45 \\
$(\varepsilon^{'}/\varepsilon )_{min}$
&--0.2 &--0.2 &--0.5 &--0.2 &--0.3 &--0.4 &--0.2 &--0.2 &--0.4 \\
$(\varepsilon^{'}/\varepsilon )_{max}$
&--0.1 &--0.1 &--0.3 &--0.1 &--0.1 &--0.2 &--0.1 &--0.1 &--0.2 \\
\hline
\end{tabular}
\\[2mm]

%=============================================================================
{\large ~~~b) Up to and including $O(p^4)$:} \\[2mm]
%
%=============================================================================
%
\small
\begin{tabular}{|c|c|c|c||c|c|c||c|c|c|}
\hline
\multicolumn{1}{|c|}{} &
\multicolumn{3}{|c||}{$\Lambda ^{(4)}_{\overline{MS}} = 215$ MeV} &
\multicolumn{3}{|c||}{$\Lambda ^{(4)}_{\overline{MS}} = 325$ MeV} &
\multicolumn{3}{|c|}{$\Lambda ^{(4)}_{\overline{MS}} = 435$ MeV\rule[-3mm]{0mm}{8mm}}
 \\ \hline
\multicolumn{1}{|c|}{$~~~~~~~~$} &
\multicolumn{1}{|c|}{$~{\bf LO}~$}  &
\multicolumn{1}{|c|}{$~{\bf NDR}~$} &
\multicolumn{1}{|c||}{$~{\bf HV}~$}  &
\multicolumn{1}{|c|}{$~{\bf LO}~$}  &
\multicolumn{1}{|c|}{$~{\bf NDR}~$} &
\multicolumn{1}{|c||}{$~{\bf HV}~$}  &
\multicolumn{1}{|c|}{$~{\bf LO}~$}  &
\multicolumn{1}{|c|}{$~{\bf NDR}~$} &
\multicolumn{1}{|c|}{$~{\bf HV}~$}\\ \hline
$\widetilde{B}_{1}$
& 4.54 & 5.13 & 4.85 & 4.16 & 4.79 & 4.42 & 3.79 & 4.41 & 3.86 \\
$\widetilde{B}_{4}$
& 0.48 & 0.43 & 0.45 & 0.51 & 0.48 & 0.47 & 0.54 & 0.46 & 0.51 \\
$P_0$
& 3.94 & 3.25 & 2.41 & 6.09 & 5.68 & 4.58 & 8.87 & 10.46& 7.87 \\
$P_2$
& 3.49 & 3.08 & 4.20 & 5.21 & 5.04 & 4.99 & 7.43 & 8.34 & 7.34 \\ 
$(\varepsilon^{'}/\varepsilon )_{min}$
& 0.4 & 0.1 &--3.1 & 0.8 & 0.6 &--0.7 & 1.2 & 1.8 & 0.5 \\
$(\varepsilon^{'}/\varepsilon )_{max}$
& 0.8 & 0.3 &--1.5 & 1.5 & 1.1 &--0.4 & 2.5 & 3.6 & 0.9 \\
\hline
\end{tabular}
\\[2mm]

%=============================================================================
{\large ~~~c) Up to and including $O(p^6)$:} \\[4mm]
%
%=============================================================================
%
\small
\begin{tabular}{|c|c|c|c||c|c|c||c|c|c|}
\hline
\multicolumn{1}{|c|}{} &
\multicolumn{3}{|c||}{$\Lambda ^{(4)}_{\overline{MS}} = 215$ MeV} &
\multicolumn{3}{|c||}{$\Lambda ^{(4)}_{\overline{MS}} = 325$ MeV} &
\multicolumn{3}{|c|}{$\Lambda ^{(4)}_{\overline{MS}} = 435$ MeV\rule[-3mm]{0mm}{8mm}}
 \\ \hline
\multicolumn{1}{|c|}{$~~~~~~~~$} &
\multicolumn{1}{|c|}{$~{\bf LO}~$}  &
\multicolumn{1}{|c|}{$~{\bf NDR}~$} &
\multicolumn{1}{|c||}{$~{\bf HV}~$}  &
\multicolumn{1}{|c|}{$~{\bf LO}~$}  &
\multicolumn{1}{|c|}{$~{\bf NDR}~$} &
\multicolumn{1}{|c||}{$~{\bf HV}~$}  &
\multicolumn{1}{|c|}{$~{\bf LO}~$}  &
\multicolumn{1}{|c|}{$~{\bf NDR}~$} &
\multicolumn{1}{|c|}{$~{\bf HV}~$}\\ \hline
$\widetilde{B}_{1}$
& 4.29 & 4.85 & 4.58 & 3.93 & 4.53 & 4.17 & 3.57 & 4.17 & 3.64 \\
$\widetilde{B}_{4}$
& 0.53 & 0.48 & 0.50 & 0.56 & 0.50 & 0.53 & 0.60 & 0.51 & 0.57 \\
$P_0$
& 3.93 & 3.23 & 2.40 & 6.08 & 5.66 & 4.56 & 8.87 & 10.44& 7.87 \\
$P_2$
& 3.57 & 3.17 & 4.29 & 5.31 & 5.15 & 5.07 & 7.55 & 8.51 & 7.43 \\
$(\varepsilon^{'}/\varepsilon )_{min}$
& 0.3 & 0.1 &--3.2 & 0.7 & 0.4 &--0.4 & 1.1 & 1.7 & 0.4 \\
$(\varepsilon^{'}/\varepsilon )_{max}$
& 0.6 & 0.1 &--1.6 & 1.3 & 0.9 &--0.8 & 2.2 & 3.3 & 0.7 \\
\hline
\end{tabular}
\newpage

%=============================================================================
%                                 Table 5
%=============================================================================
{\bf Table 5.} Upper and low bounds for $\varepsilon^{'}/\varepsilon$ 
               (in units $10^{-4}$) for different values of $\widetilde{B}_5$ 
               ($\widetilde{B}_8 = 1$) obtained by the scanning method.

\vspace{5mm}
\begin{center}
{\small
\begin{tabular}{|c|c|*{2}{c}|*2{c}|*2{c}|} \hline \hline
$~\widetilde{B}_5~$& $\Lambda ^{(4)}_{\overline{MS}},$&\multicolumn{2}{|c|}{LO}&
\multicolumn{2}{|c|}{NDR}&\multicolumn{2}{|c|}{HV} \\
\cline{3-8}
$~~~~~$& (MeV)&min&max&min&max&min&max \\
\hline
      &  215   &-3.8 & 8.5 &-4.0 & 7.5 &-7.2 & 3.3 \\
  1.0 &  325   &-4.5 &11.9 &-5.1 &11.5 &-6.6 & 8.1 \\
      &  435   &-5.5 &16.1 &-5.9 &19.6 &-6.6 &13.8 \\ \hline 
      &  215   &-2.4 &16.3 &-2.7 &14.7 &-5.6 & 9.7 \\
  1.5 &  325   &-2.6 &22.2 &-3.1 &21.9 &-4.1 &16.9 \\
      &  435   &-2.9 &29.6 &-2.9 &35.6 &-4.2 &26.7 \\ \hline 
      &  215   &-0.9 &24.0 &-1.3 &22.0 &-4.4 &16.2 \\
  2.0 &  325   &-0.7 &32.5 &-1.2 &32.3 &-2.5 &25.7 \\ 
      &  435   &-0.4 &43.2 & 0.1 &51.4 &-1.8 &39.5 \\ \hline \hline
\end{tabular}
}
\end{center}
\newpage

%=============================================================================
%                                 Table 6
%=============================================================================
{\bf Table 6.} Upper and low bounds for $\varepsilon^{'}/\varepsilon$ 
               (in units $10^{-4}$) for different values of $\widetilde{B}_5$
               ($\widetilde{B}_8 = 1$) obtained by Gaussian method. The limits
               without brackets correspond to the confidence level of 68\% 
               while the limits in brackets correspond to the confidence level
               95\%.  

\vspace{5mm}
\begin{center}
{\small
\begin{tabular}{|c|c|*{2}{c}|*2{c}|*2{c}|} \hline \hline
$~\widetilde{B}_5~$& $\Lambda ^{(4)}_{\overline{MS}},$&\multicolumn{2}{|c|}{LO}&
\multicolumn{2}{|c|}{NDR}&\multicolumn{2}{|c|}{HV} \\
\cline{3-8}
$~~~~~$&(MeV)&min&max&min&max&min&max \\
\hline
       & 215 & -2.1  & 3.0  &  -2.4 & 2.4  &  -4.9 & -0.2  \\
       &     &( -5.1 & 7.2 )&( -5.1 & 6.4 )&( -8.0 & 2.6 ) \\ \cline{2-8}
  1.0  & 325 & -2.3  & 4.3  &  -2.7 & 4.0  &  -3.6 & 2.1   \\
       &     &( -6.3 & 9.9 )&( -6.6 & 9.7 )&( -7.6 & 6.5 ) \\ \cline{2-8}
       & 435 &  -2.6 & 5.9  &  -2.5 & 7.6  &  -3.5 & 4.7   \\
       &     &( -7.8 &13.4 )&( -8.1 &16.6 )&( -8.5 &11.7 ) \\ \hline \hline
       & 215 &  -0.5 & 7.3  &  -0.8 & 6.5  &  -3.2 & 3.1   \\
       &     &( -3.8 &14.4 )&( -3.9 &13.0 )&( -6.7 & 8.3 ) \\ \cline{2-8}
  1.5  & 325 &  -0.2 &10.1  &  -0.5 & 9.8  &  -1.5 & 6.9   \\
       &     &( -4.6 &19.5 )&( -4.9 &19.3 )&( -5.9 &14.4 ) \\ \cline{2-8}
       & 435 &  -3.8 &14.4  &  -3.9 &13.0  &  -6.7 & 8.3   \\ 
       &     &( -5.5 &25.8 )&( -5.7 &30.8 )&( -6.3 &23.6 ) \\ \hline \hline
       & 215 &   0.8 &11.8  &   0.5 &10.6  &  -1.8 & 6.6   \\
       &     &( -2.9 &21.8 )&( -3.1 &20.0 )&( -5.7 &14.4 ) \\ \cline{2-8}
  2.0  & 325 &   1.5 &16.0  &   1.2 &15.8  &   0.1 &11.9   \\ 
       &     &( -3.4 &29.1 )&( -3.8 &28.9 )&( -4.7 &22.8 ) \\ \cline{2-8}
       & 435 &   2.4 &21.0  &   3.2 &24.9  &   1.4 &19.1   \\
       &     &( -4.0 &36.5 )&( -4.0 &40.7 )&( -4.8 &34.5 ) \\ \hline \hline
\end{tabular}
}
\end{center}

\end{document}